# *Updating Quantum Cryptography Report ver. 1 (May 2009)*

**Contributors**

Donna Dodson (NIST), Mikio Fujiwara (NICT), Philippe Grangier (CNRS), Masahito Hayashi (Tohoku University), Kentaro Imafuku (AIST), Ken-ichi Kitayama (Osaka University), Prem Kumar (Northwestern University), Christian Kurtsiefer (National University of Singapore), Gaby Lenhart (ETSI), Norbert Lütkenhaus (University of Waterloo), Tsutomu Matsumoto (Yokohama National University), William J. Munro (Hewlett-Packard Laboratories), Tsuyoshi Nishioka (Mitsubishi Electric), Momtchil Peev (Austrian Research Center), Masahide Sasaki (NICT), Yutaka Sata (Toshiba Research Europe), Atsushi Takada (NTT), Masahiro Takeoka (NICT), Kiyoshi Tamaki (NTT), Hidema Tanaka (NICT), Yasuhiro Tokura (NTT), Akihisa Tomita (NEC), Morio Toyoshima (NICT), Rodney van Meter (Keio University), Atsuhiro Yamagishi (IPA), Yoshihisa Yamamoto (Stanford University), Akihiro Yamamura (Akita University)

Contact address: psasaki-at-nict.go.jp <Masahide Sasaki (NICT)>

## UQC Steering Committee

- Mikio Fujiwara (NICT)
- Kentaro Imafuku (AIST)
- Mitsuru Matsui (Mitsubishi Electric)
- Masahide Sasaki (NICT)
- Yutaka Sata (Toshiba Research Europe)
- Hidema Tanaka (NICT)
- Yasuhiro Tokura (NTT)
- Akihisa Tomita (NEC)
- Atsuhiro Yamagishi (IPA)

## UQC Working Group

- Donna Dodson (NIST)
- Mikio Fujiwara (NICT)
- Philippe Grangier (CNRS)
- Masahito Hayashi (Tohoku University)
- Kentaro Imafuku (AIST), Co-chair
- Ken-ichi Kitayama (Osaka University)
- Prem Kumar (Northwestern University)
- Christian Kurtsiefer (National University of Singapore)
- Gaby Lenhart (ETSI)
- Norbert Lütkenhaus (University of Waterloo)
- Tsutomu Matsumoto (Yokohama National University)
- William J. Munro (Hewlett-Packard Laboratories)
- Tsuyoshi Nishioka (Mitsubishi Electric)
- Momtchil Peev (Austrian Research Center)
- Masahide Sasaki (NICT), Chair
- Yutaka Sata (Toshiba Research Europe)
- Masahiro Takeoka (NICT)
- Kiyoshi Tamaki (NTT)
- Hidema Tanaka (NICT)
- Yasuhiro Tokura (NTT)
- Akihisa Tomita (NEC)
- Atsuhiro Yamagishi (IPA)
- Akihiro Yamamura (Akita University)

# Table of Contents

# About this Report

Quantum cryptographic technology (QCT) is expected to be a fundamental technology for realizing long-term information security even against as-yet-unknown future technologies. More advanced security could be achieved using QCT together with contemporary cryptographic technologies. To develop and spread the use of QCT, it is necessary to standardize devices, protocols, and security requirements and thus enable interoperability in a multi-vendor, multi-network, and multi-service environment. This report is a technical summary of QCT and related topics from the viewpoints of

1. *consensual establishment of specifications and requirements of QCT for standardization and commercialization and*
2. *the promotion of research and design to realize New-Generation Quantum Cryptography.*

Here, QCT includes cryptographic protocols and related technologies that use the laws of quantum mechanics.

This report comprises four parts.

**Part I:** *Status and Trends of QCT and Related Topics*, including a tutorial for those not familiar with concepts used in the field.
**Part II:** *Consensual Building of Specifications and Requirements of QCT for Standardization and Commercialization,* providing an organizational plan for basic concepts as well as challenging issues.
**Part III:** *Toward New Generation Quantum Cryptography,* addressing how to achieve innovation in information and communications technology by combining QCT with photonic network technology, and presenting challenging issues and effective crosscutting approaches.
**Part IV:** *Summary*, including an implementation plan and organization proposal.

We make three additional comments about the report.

1. The Updating Quantum Cryptography Working Group (UQCWG) continues to edit/revise this report, basically in accordance with the international conference of Updating Quantum Cryptography.
2. The UQCWG makes an effort to reflect a wide range of expert advice in the report, and will make the report available to the public when completed (not necessarily in a perfect form).

3. The contents of this report can be used or cited either partially or in full in any other proposal documents by public organizations/projects in countries of the contributors. The report may be translated for such purposes if the UQCWG is properly cited. The copyright belongs to the UQC steering committee.

# Part I: Status and Trends of QCT

## I – 1 Concepts and Principles

Contemporary cryptographic technologies rely on huge computational cost or the presumed difficulty of mathematical problems. Their security is guaranteed only by algorithms, which are almost always used in the application layer. Security is always threatened by new technologies such as the increasing power of computers, new mathematical algorithms, and new cryptanalysis techniques. For example, the current RSA1024 scheme will cease to work around the year 2015, when super computers will be able to factorize composite numbers of 1024 bits into their primes. The current scheme needs to be updated to use doubled key-length (RSA2048), which would indeed be a tedious task for various systems.

Quantum cryptography provides security using the laws of quantum mechanics. Currently, several types of protocols of quantum key distribution (QKD) and quantum-noise randomized cipher (QNRC) have been established. Some QKD protocols have been certified by proofs of unconditional security; specifically, information-theoretic security protocols have been confirmed to be resistant to any possible attack. In information-theoretic security, the amount of information leaked to an eavesdropper can be made arbitrarily small as the key length or code length is increased. No future technology can break such security. Such protocols provide long-term security and especially conform to the demands of national information security, secure and confidential communication by financial entities, and any mission-critical applications for which a security breach of encoded data could be a disaster not covered by an insurance mechanism.

Other QKD protocols are still under investigation in terms of proof of their complete security. They are, however, known to be already secure against very advanced technologies that will not be realized for a few decades yet.

Quantum cryptography uses optical communication channels that are governed by the laws of quantum mechanics. The practical security level directly depends on the physical conditions of implementation. This is in sharp contrast to contemporary cryptography, for which the security level is merely determined by an algorithm. This particular feature of quantum cryptography requires certification methods that are somewhat different from those for conventional cryptoschemes in providing users with practical solutions. In quantum cryptography, there are generally trade-offs between the practical security level and implementation cost. Analysis of quantum cryptography is more involved and must be a central issue in the certification and standardization of quantum cryptography.

## 1. QKD

QKD protocols are schemes that expand a secret key among two players while maintaining information-theoretic security for the key. These protocols enable the two players to terminate their protocols when they detect intrusion using the quantum uncertainty principle. Such an extremely secure key could be used in various data encryptions, but the most attractive scenario at present is to realize information-theoretic secure private communications between two players applying the secret key to a one-time pad (OTP) scheme, which requires entropy of the secret key as much as that of the message.

To implement QKD protocols in practice, we need devices that are not familiar to current optical networks such as single-photon detectors and sources of extremely weak coherent pulses (WCPs) and entangled photons. In addition, the security of such protocols is guaranteed by information-theoretic security based on physical (quantum mechanical) properties of the channel. These protocols are very different from contemporary cryptographic protocols based on complexity-theoretic security and mathematical algorithms. While these processes result in the superiority and uniqueness of QKD, one has to newly formalize its concepts and specifications from both the viewpoints of security and device requirements.

## 2. QNRC

A QNRC is a physical-layer data encryption scheme that uses quantum noise to randomize the observed output of a cipher. The goal of implementing a QNRC is to realize a practical random cipher system. A random cipher is an extended version of a standard cipher where cipher text is further randomized by a random variable that is unknown to both a receiver, Bob, and an eavesdropper, Eve. In contrast to a standard random cipher for which a sender, Alice, generates random numbers artificially, a QNRC utilizes quantum noise as an automatic source of random numbers. The latter is practically attractive since it is a truly physical randomness and the transmission can be as fast as the communications system (i.e., not limited by a random-number generation rate). The noise produced using this technique is not necessarily quantum mechanical, and thus the protocol itself has a wider flexibility in its implementation. Indeed, the required physical technologies for a typical protocol of a QNRC, called Y00 or $\alpha\eta$, are almost compatible with current optical network infrastructures.

Quantitative security analysis of QNRCs is, however, still under investigation. In practice, it is expected that the additional randomness provides complexity-theoretic security better than that of conventional ciphers. Therefore, we can almost use the security notions of conventional ciphers for those of QNRCs.

It should be emphasized that a QNRC is a data encryption protocol (that uses a seed key) while QKD is a protocol to distribute a key itself. Therefore, QKD and QNRCs are not competing technologies and are being pursued independently.

## I – 2 Achievements

### I – 2 – 1 Theories

#### 1. QKD protocols and their security proofs

Inspired by an idea of Wiesner, the first QKD protocol was proposed and published in 1984 by Bennett and Brassard [Bennett and Brassard, 1984]. This protocol is designed to securely distribute a secret key between two parties, and this secret key is to be used in OTP encryption [Vernam, 1926]. Thus, using this QKD protocol (BB84), we expect the realization of very secure communications. BB84 uses four single-photon polarization states belonging to two conjugate bases, say the $Z$ and $Y$ bases for spin. Since the proposal of BB84, many QKD protocols have been proposed.

In 1991, Ekert considered a protocol for which the security is based on a violation of Bell's inequality [Ekert, 1991]. In this protocol (E91), each of two legitimate parties share a part of an entangled photon-pair and makes a measurement on each photon with bases randomly chosen from three bases. The intrinsic quantum correlation of the photon-pair allows the two parties to share the key. The basic idea for preventing eavesdropping is as follows. Imagine that Eve makes a measurement of the particles and resends states. This eavesdropping introduces the so-called physical reality to the spin measurement and this reality is detected via a test based on Bell's inequality. In 1992, Bennett, Brassard, and Mermin applied the concept of E91 to BB84. Each party performs the same measurement as Bob does in BB84; i.e., each party randomly chooses the $Z$ or $Y$ basis for the measurement [Bennett et al., 1992b]. This protocol is referred to as BBM92, and both E91 and BBM92 rely on entanglement for their security.

Also in 1992, Bennett proposed another important protocol, referred to as B92 [Bennett, 1992]. Unlike E91 and BBM92, this protocol does not directly use entanglement as the security basis. Since B92 uses only two nonorthogonal states, it is the simplest protocol. Note that any two nonorthogonal states can be unambiguously discriminated with some probability. It follows that the use of single-photon polarization as the encoding media cannot achieve long distances since Eve can exploit channel losses in such a way that she sends the vacuum state when she fails to identify the states. Thus, in order to cover long distances, we need an auxiliary system that detects the suppression of signals. For this purpose, Bennett proposed the use of strong reference light together with two signal lights.

BB84 and B92 seem to suggest that the use of nonorthogonal states is the essence of the security; however, this is not always the case. Goldenberg and Vaidman proposed a protocol that uses two orthogonal states [Goldenberg and Vaidman, 1995]. This state is a two-mode state, and the point is that Alice sends the signal in such a way that Eve cannot have access simultaneously to both modes. In this manner, Alice prevents Eve from free access to the information.

With the increase in interest in QKD, researchers have attempted to implement QKD protocols in practice. Because of its relative simplicity, BB84 was widely demonstrated experimentally. A major problem in the implementation of BB84 is how one prepares a single-photon state. In most experiments, an attenuated coherent light source is used instead of a perfect single-photon source while some works have been devoted to the creation of a perfect single-photon source. However, all photon sources so far have some probability of multi-photon emission, from which Eve can obtain information freely by exploiting the so-called photon number splitting (PNS) attack [Brassard et al., 2000]. Since this attack is one of the greatest threats to BB84, protocols with PNS tolerance have been considered. The differential phase shift (DPS)-QKD [Inoue et al., 2002; Inoue et al., 2003], SARG04 protocol [Scarani and Aćin, 2004], and decoy state method [Hwang, 2003; Lo et al., 2005; Wang, 2005] are examples of such protocols or PNS-attack-resistant methods.

What we described in the previous paragraph is a good example of how theorists and experimentalists collaborate, and it also suggests that it is very important to consider particularly threatening attacks. On the other hand, to analyze whether a QKD protocol is secure against any eavesdropping, i.e., unconditionally secure, is also very important task in the theoretical research in QKD. The first unconditional security proof for BB84 was presented by Mayers in 1996 [Mayers, 1996]. In the proof, Alice is assumed to have a perfect single-photon source while Bob has threshold detectors. This proof was generalized by Inamori, Lütkenhaus, and Mayers [Inamori et al., 2001] to accommodate the use of a coherent light source. Since these security proofs provide insights into security in a complicated manner, works have been devoted to simplifying these insights. One successful simple security proof is based on the distillation of a maximally entangled state (entanglement distillation protocol), which was proposed by Lo and Chau in 1998 [Lo and Chau, 1998] and Shor and Preskill in 2000 [Shor and Preskill, 2000]. The proof by Lo and Chau assumed possession of a quantum computer, and this assumption was later removed in the proof by Shor and Preskill, who assumed a perfect single-photon source and photon-number-resolving detectors. Because of the simplicity of the proof based on the entanglement distillation protocol, the proof was generalized, mainly in three directions. First, the achievable distance of secure communication, for instance, was increased by considering two-way classical communications [Gottesman and Lo, 2003]. Second, imperfections such as multi-photon emissions and the use of a threshold detector were taken into account [Koashi and Preskill, 2003; Gottesman et al., 2004; Tsurumaru and Tamaki, 2008; Koashi et al., 2008; Beaudry et al., 2008]. Third, the proof was applied to other protocols such as B92 [Tamaki et al., 2003; Tamaki and Lütkenhaus, 2004; Koashi, 2004; Tamaki et al., 2006], the three-state protocol [Phoenix et al., 2000; Boileau et al., 2005], and SARG04 [Tamaki and Lo, 2006]. We note that all these generalizations are important since they are useful in increasing the communication distance, thus making QKD more practical, and indicating which protocol should be used on the basis of experimental parameters. In 2005, the proof was further simplified by Koashi, who considered the distillation of the eigenstate of a qubit instead of the maximally entangled state [Koashi, 2005]. Though we do not know yet whether this proof can accommodate the use of two-way classical communication, the proof has advantages in terms of the bit error rate threshold and simplicity in treating the imperfections of devices.

Another important security proof that is not based on the distillation of a pure state but on an information-theoretic concept was presented by Kraus, Gisin, and Renner [Kraus et al., 2005]. This security proof can also be applied to protocols such as B92 and SARG04. One of the striking features of this proof is that it rigorously shows that if Bob applies appropriate noise intentionally to the data (a process referred to as post-processing), then the bit error rate threshold increases. This is the first formal proof of an intuition that in some situations adding noise on Bob's side may disturb more correlation between Eve and Bob than the one between Alice and Bob, which is advantageous to Alice and Bob.

Finally, Horodecki et al. [Horodecki et al., 2005] showed that QKD can be based on underlying entanglement concepts without the actual distilling of the entanglement, and thus the secret key can also be obtained from the so-called bound entangled states, which cannot be distilled in principle.

So far, we have discussed only discrete variable QKD where a single-photon level detection is required. This detection is technologically difficult, and a scheme involving the measurement of the continuous data, i.e., amplitude of the light, by using homodyne detection was proposed by Ralph in 1999 [Ralph, 1999], Hillery in 2000 [Hillery, 2000], and Reid in 2000 [Reid, 2000]. These protocols and their generalized version are called continuous variable (CV) QKD since Bob's raw data is continuous. In these protocols, Alice encodes her bit data by performing phase or amplitude displacement of a light pulse. When the phase-space displacement values (in the complex plane) are discrete, this family of CV protocols is called CV QKD with discrete modulation. On the other hand, there is another family of CV QKDs, the CV QKD with Gaussian modulation where the signal amplitude is modulated according to Gaussian distribution in quadratures $x$ or $p$. This kind of QKD was proposed by Cerf, Lévy, and Van Assche in 2001 [Cerf et. al., 2001], using squeezed states of light, and by Grosshans and Grangier in 2002 [Grosshans and Grangier, 2002], using coherent states.

The first experimental demonstration of CVQKD was carried out by Grosshans et al in 2003 [Grosshans et al., 2003], using an improved protocol called "reverse reconciliation", which allows in principle to distribute a key through a line with arbitrary losses. It was shown also that CV QKD may involve either homodyne detection or heterodyne detection, which makes it fast and efficient [Lance et al., 2005]. These features are considered to be one of the advantages over discrete variable QKD. Examples of up-to-date CVQKD setups with Gaussian modulation, able to distribute keys in an automated way within a secure network, are given in [Lodewyck et al., 2007] and [Fossier et al., 2008]. As pointed out by Silberhorn et al [Silberhorn et al., 2002], the discrete modulation schemes can operate also beyond 3dB channel loss, and by a combination of post-selection mechanisms and practical one- or two-way error correction they can be made stable for real implementations.

Although practical CV QKD is relatively newly proposed, its security has rapidly been investigated. Many works have been devoted aiming to prove the unconditional security of CV QKD. The first practical protocol [Grosshans et al., 2003] was proven secure against individual Gaussian attacks. The proof was quickly extended to arbitrary individual attacks [Grosshans and Cerf, 2004], and then to arbitrary collective attacks [Navasques et al., 2006; Garcıa-Patron and Cerf, 2006]. Novel protocols, designed to increase the transmission distance, were also proposed for example by Leverrier and Grangier in 2008 [Leverrier and Grangier, 2008] or Zhao, Han, and Guo in 2008 [Zhao et al., 2008].

Finally, it was proven by Renner and Cirac in 2009 [Renner and Cirac, 2009] that coherent attacks are in the asymptotic limit not more powerful than collective attacks, establishing the unconditional security of CVQKD protocols based on the Gaussian optimality theorem [Cirac et al., 2006]. It should be pointed out that this theorem is not applicable to the protocols using post-selection, which are thus not proven fully secure so far.

As we have seen, the theory of QKD is rapidly growing area, and it is not only about making new proposals to experimentalists but also receiving feedbacks from them in order to give back them new proposals to relax the constraints of devices or to enhance the achievable distances. Thus, by interacting each other, experimentalists and theorists are working together to close the gap between theory and experiment.

## 2. QNRC

The first discussion on the possibility of data encryption via quantum noise randomization was initiated by a protocol proposed by Yuen in 2000 [Yuen, 2003] called Y00 or $\alpha\eta$. In addition to Y00 being the first data encryption protocol based on quantum mechanics, it is attractive from a practical viewpoint since it uses relatively strong coherent states (more than a single-photon level) and a homodyne detector, which is faster and more efficient than photon counters.

After discussions on its security issues, nowadays the Y00-based data encryption protocol is recognized as a QNRC protocol [Nair et al., 2006]. A QNRC is a physical layer data encryption that uses quantum noise to randomize the observed output of a cipher. It is based on the (classical) random cipher. In contrast to standard ciphers, random ciphers use an additional random variable $R$ in addition to an initially shared key, where $R$ is generated by Alice, known to only her, and used in data encryption. The randomization is done such that Bob can decrypt his received cipher using only the shared key, while it acts as inevitable noise for Eve. This additional $R$ increases the security of ciphers (in practice, computational cost required for Eve). However, the implementation of the proposed random ciphers faces a large obstacle owing to a reduction in the data rate. Moreover, a true and high-speed random number generator (RNG) is hardly obtainable at present.

A QNRC provides a practical high-speed (in excess of gigabits per second) random cipher in the physical layer. In Y00 and related protocols, the additional randomness is effectively provided by quantum noise in an optical pulse instead of by Alice's RNG. Due to the law of quantum mechanics, quantum noise cannot be perfectly removed. Thus, properly choosing the physical signal modulation format, quantum noise acts as the random variable $R$. Compared with using artificial RNGs, quantum noise has two advantages: it provides truly physical randomness and it realizes high-speed operation. (Since quantum noise is automatically embedded in every mode of optical signals, the data rate is not limited by the speed of random number generation.) In addition, the physical system and devices required for the implementation of Y00 are almost compatible with current optical network infrastructure and devices. In the total system, quantum-noise-based randomization is combined with contemporary data encryptions such as the advanced encryption standard (AES). Therefore, quantum noise randomized symmetric ciphers (QNRSCs) might be practical.

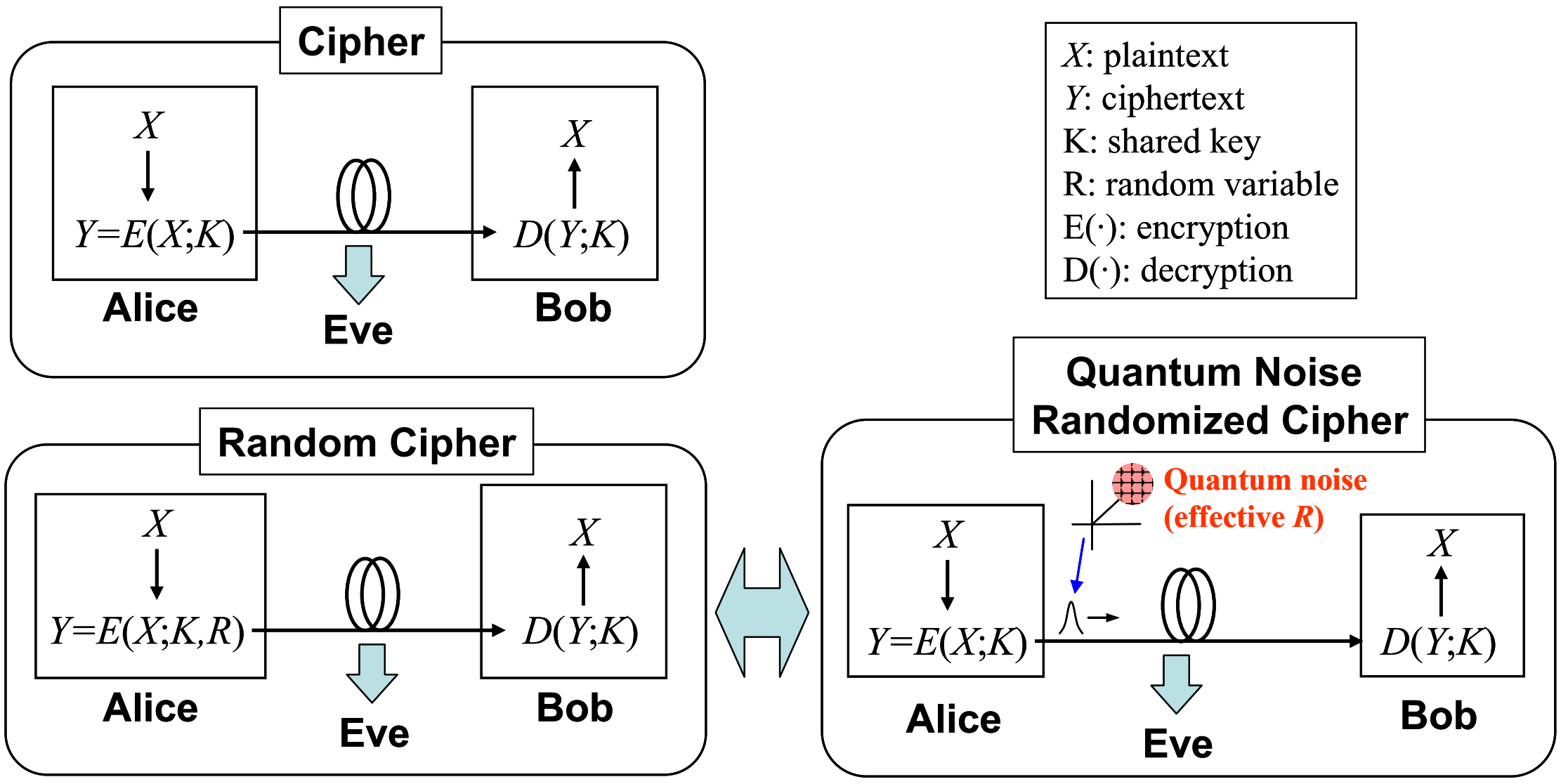

Quantitative security analyses of QNRSCs have not yet produced clear results. Obviously, the security of a QNRSC protocol is equivalent to that of a certain class of classical random symmetric cipher. Therefore, the security analysis can be mostly conducted using the theory of contemporary ciphers, and it is highly desirable to establish the practical theory clarifying how large the difference in security is between random and nonrandom symmetric ciphers. Once such theory is developed, one could be estimate the security of QNRSCs quantitatively by incorporating Eve's physical attacks based on quantum mechanical processes.

### I – 2 – 2 Experiments

This section briefly reviews progress on the experimental realization of quantum cryptography. The present review does not provide a complete list of experimental achievements, but traces steps made toward the realization of practical quantum cryptographic systems.

### 1. QKD

**Steps toward practical QKD systems**

Experimental studies in QKD is typically advanced by taking a few conceptual steps. First, proof-of-principle experiments show that the signals can really be transmitted while maintaining their quantum mechanical characteristics. In QKD experiments for example, interference or successful transmission of a coherent superposition of photons is demonstrated. For this purpose, it suffices to transmit sequences of identical fixed states. In a next step, key transmission first with a fixed pattern, and then with a randomly selected sequence is performed to include dynamical effects of the system. The transmission and detection of signals is, however, only half of the task; the transmission needs to be followed by post-processing. Post-processing guarantees the security of the implemented protocol, which should be supported by an appropriate theory with reasonable assumptions. This sequence of steps describes the historical development of QKD, but it is at the same time the meaningful path of steps in implementing and testing novel QKD implementations.

Along with the progress in key transmission and post processing, the system control needs to be integrated with some steps for the QKD experiment to turn into a QKD application. First, the transmitter and receiver may work with a clock signal from the same clock oscillator, typically meaning that sender and receiver have to sit in a laboratory side to side. Then, isolated clock generators need to be implemented, along with a proper clock synchronization function. To obtain a correct key in practical system, real-time frame synchronization should be also implemented. A proto-type of the QKD system will then also require a fault detection-recovery mechanism. The proto-type will be sophisticated to a commercial point-to-point QKD system, when for the classical post-processing protocols are protected by a properly implemented authentication function.

**Transmission and post-processing**

The QKD protocol was first demonstrated in 1992 [Bennett et al., 1992a]. The transmission distance was only 30 cm. Since then, experimental research has mainly focused on improving transmission performances, such as increasing the communication distance and key-generation rate. The first step to long-distance QKD transmission might be the use of optical fibers. It was not clear whether an optical fiber conserves the coherence of a single photon after transmission of several tens of kilometers in the early state. Muller et al. performed a remarkable demonstration in which the polarization state of a single photon was conserved through an optical fiber cable installed under a lake [Muller et al., 1996]. Since polarization states are easily altered by strain acting on optical fibers, phase coding has been proposed. Single-photon transmission through a 10 km optical fiber was demonstrated [Townsend, 1994]. Currently, QKD transmission through a 202 km optical fiber (the effective length is shorter: 148.7 km for $\mu = 0.1$ and 184.6 km for $\mu = 0.5$) has been reported using an extremely low dark noise photon detector [Hiskett et al., 2006]. Recently, long-distance QKD experiments have been reported using installed fiber links: 67 km transmission in Switzerland in 2002 [Stucki et al., 2002], 96 km in Japan in 2005 [Hasegawa et al., 2005], and 125 km in China in 2005 [Mo et al., 2005].

Along with the long distance transmission, the clock frequency has been increased for high-speed key generation. In early experiments, the clock frequency was about 1 MHz owing to the afterpulse effect in avalanche photodiode (APD) photon detectors. Recently, a 625 MHz clock has been used in a QKD system with fast superconducting single-photon detectors (SSPDs) [Tanaka et al., 2008]. If the photon detection rate is not too high, an APD-based QKD system can also operate with a gigahertz clock frequency [Yuan et al., 2008]. Another figure of merit of the single-photon detector is the timing jitter, which is especially important for a very high clock frequency system (~10 GHz). SSPDs show superior timing jitter to other single photon detectors.

Free-space QKD systems have also been improved in transmission performances. One of the main targets of the free-space QKD system for free-space transmission is to construct an Earth–satellite link. The system should support high-speed transmission over a long distance between the earth and a satellite, since allowable communication time will be only a few minutes. In free-space systems, bits are encoded in polarization states, which are conserved in the air. The atmospheric transmission window enables us to use visible light, and thus a high-performance Si-APD can be applied to free-space QKD systems. QKD transmission was demonstrated even in daytime in 2000 [Buttler et al., 2000]. A QKD demonstration was successful over 144 km [Schmitt-Manderbach et al., 2007]. Operations with high clock frequencies have also been investigated in the gigahertz range [Bienfang et al., 2004].

Experimental studies have focused on generating keys with low quantum bit error rates (QBERs) and examined the security of the generated key with the criteria derived for ideal systems. The security in real systems has been considered only recently. A decoy method has been implemented to increase the final key rate [Zhao et al., 2006; Peng et al., 2007; Yuan et al., 2007; Rosenberg et al., 2007] of the system using a dimmed laser as a light source. Recently, theoretical investigations were conducted to estimate the effect of statistical variation in finite-length data. It has been shown that careful estimation of the number of sacrifice bits is necessary to guarantee the security of the finite-length code. Secure key distillation has been implemented by taking account of the effect of the finite-length code [Hasegawa et al., 2007; Scarani and Renner, 2008].

The above QKD systems are intended to implement the BB84 protocol. High transmission performances have been reported in DPS-QKD systems [Diamanti et al., 2006; Takesue et al., 2007] that employ either a 1 GHz clock frequency with a frequency up-conversion single-photon detector or a 10 GHz clock with a SSPD.

Entanglement-based QKD systems have been implemented for fiber transmission [Jennewein et al., 2000; Tittel et al., 2000] and free-space transmission [Naik et al., 2000; Ursin et al., 2007; Erven et al., 2008]. Recently, a modified E91 protocol has been demonstrated, where the information obtained by the eavesdropper was estimated from the violation of the CHSH Bell equation [Ling et al., 2008].

A CV-QKD system consisting of a coherent state with Gaussian modulation and homodyne detection has also been investigated for fiber channels. Recently, a higher generation rate of the secret key over a relatively short distance (1.5 kbps, 25 km) has been achieved [Lodewyck et al., 2007], with the generation rate being limited simply by the computer speed. Implementation of a plug-and-play CV-QKD with discrete modulation and homodyne detection has also been examined [Hirano et al., 2006].

**Implementation, system control, and security**

Phase coding BB84 systems require two asymmetric Mach–Zehnder interferometers to create a coherent double-pulse for encoding and to generate interference between the double-pulse for decoding. The two interferometers should have identical path differences to obtain interference with high visibility. A plug-and-play system solves this issue using the same interferometer for encoding and decoding in the round-trip architecture [Muller et al., 1997]. The plug-and-play system provides stable operation using conventional optical devices without precise control of the interferometer. A number of QKD systems including commercial systems (idQuantique and MagiQ) employ the plug-and-play architecture. However, the plug-and-play systems suffer from back scattering light owing to their round-trip architecture. Burst operation can reduce the back scattering noise with the cost of decreasing the operation throughput to one-third, which limits the high-speed key generation. It has been pointed that the plug-and-play system is vulnerable to Trojan-horse attack. This system also reveals a phase reference, which may allow more efficient eavesdropping. Therefore, one-way architecture is revived with stabilizing techniques. One technique is the active control of the interferometer [Yuan and Shields, 2005]; another is the use of planar lightwave circuits (PLCs) [Nambu et al., 2004]. A PLC is an integrated optical circuit that is mechanically stable with a small footprint. The interference can be maintained only with temperature control. A DPS-QKD system simplifies the requirement of the interferometer; only one interferometer is needed in the decoder, the path difference of which is adjusted to the period of the coherent optical pulse train. A PLC is also useful in DPS-QKD systems. One of the important aspects of security is to avoid using active devices since they may cause security loopholes that can be used by Trojan-horse attack. The simple setup of a DPS-QKD system using only passive devices might be another merit.

Currently, a single-photon source is becoming a less crucial component of a QKD system; however, it is still useful in simplifying the protocol. A QKD system with a single-photon source operating at 1310 nm has been demonstrated [Intallura et al., 2007]. A single-photon source operating at 1550 nm has also been reported [Miyazaki et al., 2005]. An alternative device, the heralded single-photon source, was used for QKD transmission [Soujaeff et al., 2007].

A practical communication system should contain function blocks for the clock synchronization, frame synchronization, and fault detection/recovery to ensure stable operation. A system equipped with the above functions provided continuous hands-free operation over 14 days through an aerially installed optical fiber link. Continuous operation was also reported [Tanaka et al., 2005; Tajima et al., 2007]. Recently, the Development of a Global Network for Secure Communication based on Quantum Cryptography (SECOQC) project demonstrated the continuous operation of QKD systems connected with an optical fiber network in Vienna [Poppe et al., 2008].

### 2. QNRC

Experiments were performed with the intention of implementing the proposed protocol with multi-ary phase [Barbosa et al., 2003] and amplitude modulations [Hirota et al., 2005], and successful encoding–decoding was reported. Since the device and system requirements of these protocols are almost compatible with (or slightly severer than) current photonic network technologies, some prototypes for commercialization have already been successfully demonstrated; e.g., by NuCrypt (phase shift keying data transmission, 2.5 Gbps, 210 km) [NuCrypt, 2007] and Hitachi Info & Communication Engineering (amplitude shift keying video transmission, 2.5 Gbps, 50 km) [Hitachi, 2007].

## I – 3 Examples of Practical Applications and Commercial Products

Quantum cryptography is no longer a subject reserved to scientists, but is becoming a practical solution for secure communications in the real world. There are already black-box commercial products available that are easy to use. Indeed, QKD has been applied by nonspecialists to secure communications.

During the Swiss federal elections in the fall of 2007, the information technology department of the Geneva government used id Quantique's Cerberis encryption system with QKD to secure the network processing of voting results. The system was used to secure a gigabit Ethernet link connecting the central counting station located in downtown Geneva and the data center where all the results were stored and processed (see map below). The system ran perfectly during the elections. This was the world's first application of quantum cryptography. Following this successful pilot project, the chancellery of the canton of Geneva decided to acquire a Cerberis solution and rely on quantum cryptography for all future elections.

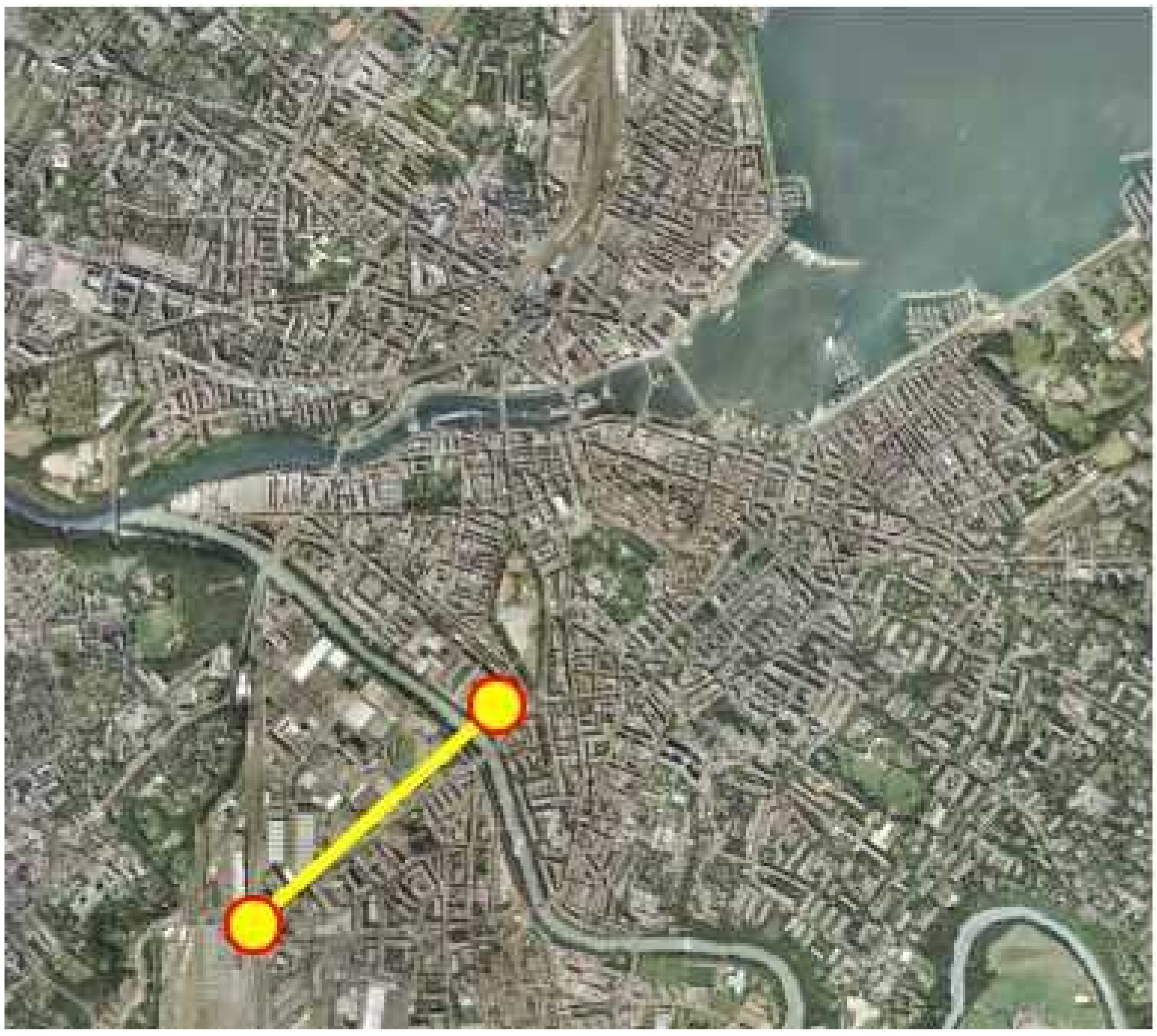

(From http://www.idquantique.com/news/news-elections2008.htm)

Some commercial QKD products have already been rolled out. Three representative products from idQunatique, MagiQ Technologies, and SmartQuantum are now introduced.

**idQuantique**

idQuantique is the first company to launch a QKD system as a commercial product. The latest product is the id3100/id3110 Clavis$^2$ QKD system, which supports BB84 and SARG04 protocols. An autocompensating interferometric plug-and-play system is adopted. Cerberis is a network security solution comprising Clasvis$^2$ and a contemporary encryption engine Centauris. The price of Clavis$^2$ is

about US$150,000–200,000.

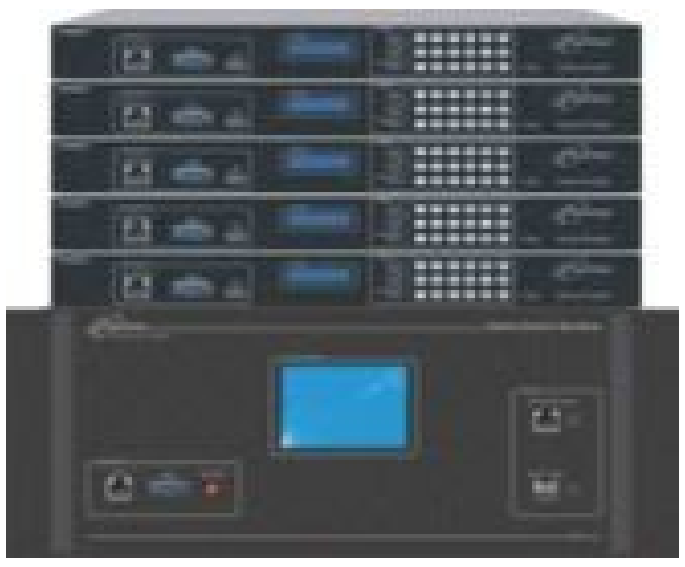

Cerberis
(From http://www.idquantique.com/)

**MagiQ Technologies**

MagiQ Technologies is another pioneer of QKD commercial products. The QKD system NAVAJO Security Gateway was launched in 2003 as a virtual private network solution. Recently, MagiQ Technologies promoted MagiQ QPN as a quantum cryptographic solution and proposed several solutions using QPN security gateways. The price of NAVAJO Security Gateway is US$200,000–400,000.

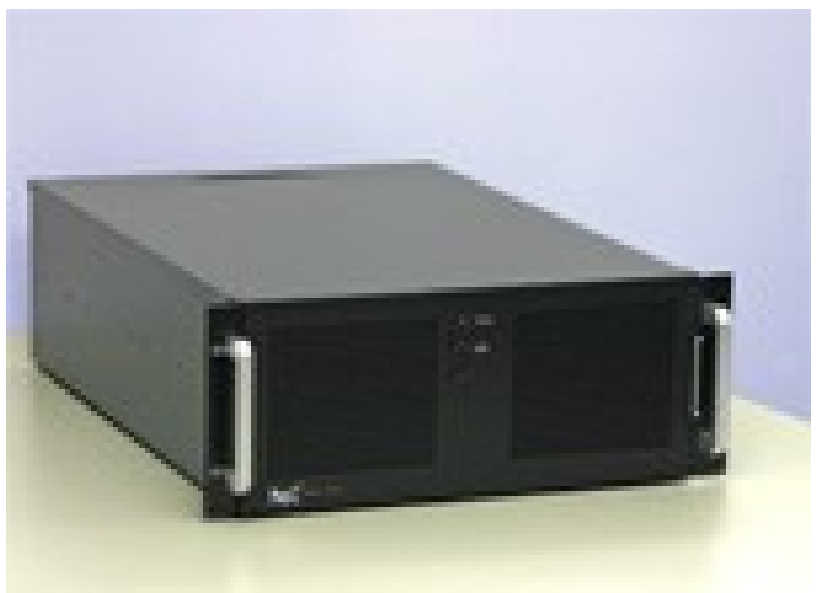

QPN Security Gateway
(From http://www.magiqtech.com/)

**SmartQuantum**

SmartQuantum also provides a QKD solution, the SQKey Generator. SQBox Defender was selected as one of the best innovative products at the Eurosatory exhibition (an international defense exhibition) in June 2008.

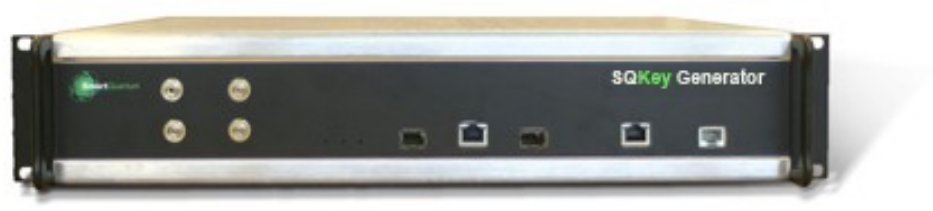

SQKey Generator

(From http://www.smartquantum.com/SmartQuantum.html)

The main specifications of the above three products are given in the following table.

Specifications of commercial products

| Product | Clavis[2] | NAVAJO Security Gateway | SQKey Generator |
|---|---|---|---|
| Vendor | idQuantique | MagiQ Technologies, Inc. | SmartQauntum Inc. |
| Protocol | BB84, SARG | BB84 | BB84 |
| Typical distance | 50 km | 50 km | 80 km |
| Key refresh rate | 1 kbps @ 25 km | 100 keys/s | one 192-bit key/s |

Combining with an OTP, these products support a hybrid system with contemporary cryptography and assure high performance compared with current optical communication. The three vendors recently promoted solutions based on a QKD system rather than only a QKD system itself.

## I – 4 Potential Applications and Markets

### 1. Physically secure private network

Many companies and mission-critical organizations have begun wanting their own private fiber network, to prevent information leaks and maintain confidentiality. Such companies want their network to be physically isolated from the public Internet, and intend to manage it themselves instead of relying on network services provided by carrier companies.

Such a network, referred to as a physically secure private network, consists of authorized terminals and a secure data center, where confidential data such as research and design data and new product design data are stored and centrally controlled. The network can be within a campus and hence be contained within a few tens of kilometers. The total key consumption per day may not be large because encryption is not always required but is used only for highly confidential data transmission and backup transmission, which may occur a limited number of times a day.

The current level of QKD would thus suffice and be an appropriate solution for secure communications in such a physically secure private network. This kind of private network could be demanded by every type of industry, social service, and business. This could act as a strong impetus to the standardization of QKD.

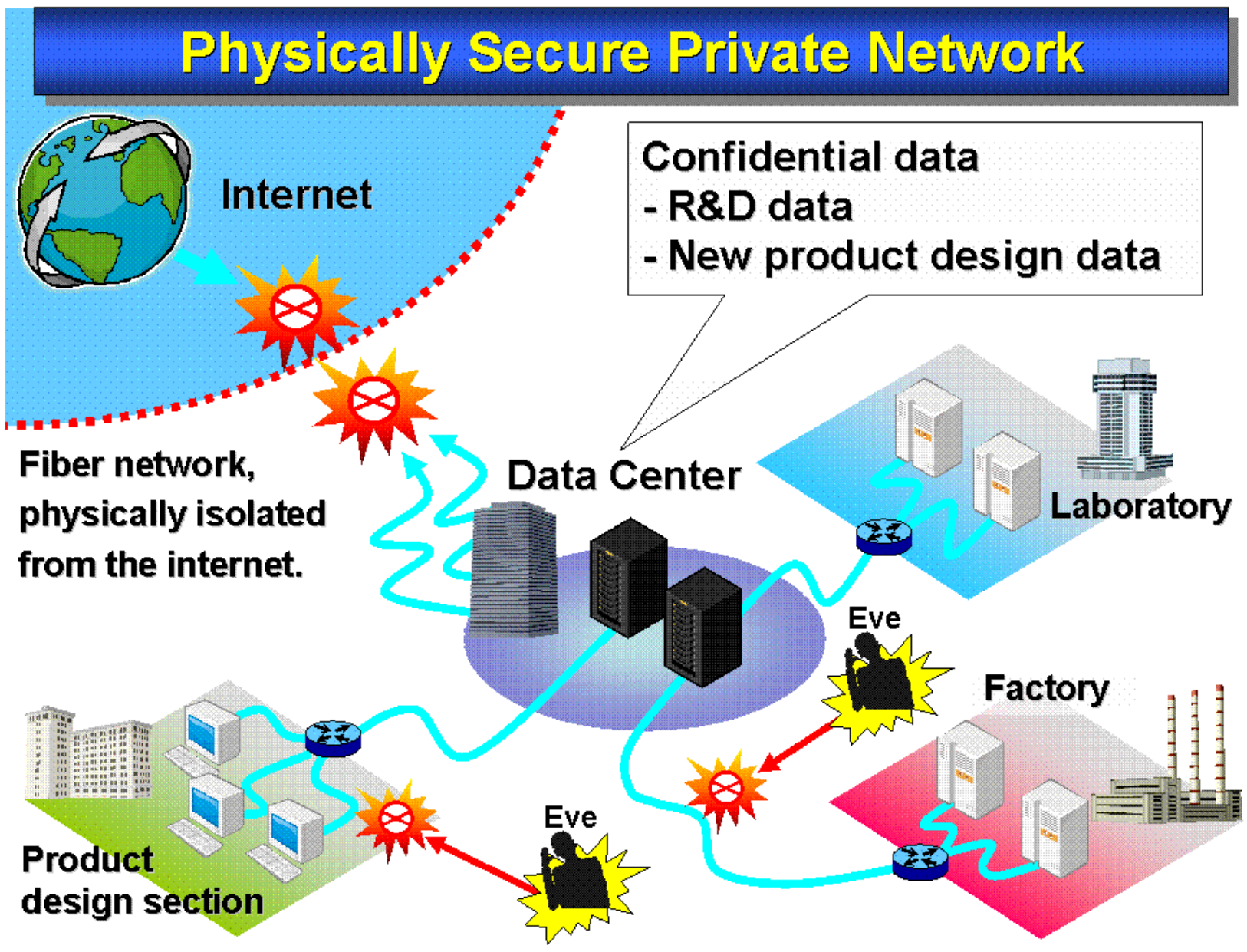

## 2. Random number sharing between mobile terminals

The security of e-commerce, and hence its user acceptance, depends critically upon asymmetric cryptography to perform user authentication and to safely transmit session keys. This form of encryption depends upon having a number of trapdoor (one-way) function operations that are prohibitively expensive for the uninformed to perform, but inexpensive for the informed. There is, however, a risk in that such systems can be broken. There is one type of cryptography that has been around for many years, the OTP, that is guaranteed to be secure whatever the advances in physics, computer science, and mathematics. The problem, then, is creating two identical copies of an OTP, with each party being given one of the copies. There are many ways of achieving this. Quantum physics provides a solution to the problem of distribution. The shared secret is generated using QKD. The shared secret is then merged into the existing identical copies of the OTP owned by both parties.

Existing QKD technology provides for secure (backbone) network infrastructure and but not for the mobile user. Current trends in consumer electronics indicate the move to small highly portable devices, for which traditional QKD infrastructure schemes will be unusable. The issue then becomes how one can securely get OTP material onto a consumer device such as a phone or personal digital assistant (PDA). Several groups have implemented a low-cost free-space short-range (less than 1 m)

QKD system between an individual and a network base station (for instance, Hewlett Packard and the University of Bristol) with the total operation time being on the order of 1 s. Such systems are based on a weak pulse BB84 implementation and require the Alice unit to be quite inexpensive (a few dollars at most, so light emitting diodes are used instead of laser diodes). In fact, the Alice units in this system are also heavily constrained in terms of available power and processing power. Bob is typically much more powerful than Alice and is a fixed infrastructure. Typically there will be many Alice units to one Bob unit.

Once it has been established securely, the OTP can be used or consumed as needed (not necessarily at the time of generation). In the classic application of the encryption of messages, the used OTP is destroyed. However, the OTP has other uses:

- Some of the OTP may be revealed, and discarded, to identify the user; e.g., a single sign on, entry to a building, and proof of identity to a bank. This revelation may be performed using a system such as Bluetooth. Some of the OTP could be used as a one-time pin number for e-commerce on the Internet. This does not require quantum technology at all. The generation of the OTP and its consumption are completely independent.
- Documents may be bulk-encrypted using an AES, or equivalent, and the key is obtained by consuming some of the OTP.
- The user might have several OTPs stored in their PDA. This is equivalent to having multiple visas in a passport. They reveal content from the appropriate visa/pad. If they accidentally reveal data from another pad, then this is not a problem because the data are simply random bits.

In terms of the OTP economy, the OTP periodically gets low. The user goes up to a Bob device and uses QKD to top up their personal OTP, one copy of which is stored in the PDA or mobile phone. This topping up process is protected using some of the existing OTP. Once topped up, daily activities are performed periodically consuming the OTP. It should also be noted that shim attacks are detected using QKD.

Finally, it is difficult to determine the market size for consumer-based QKD, but it is potentially large given the number of cell phones and PDAs worldwide, especially service-based models.

## 3. Future vision: a secure photonic network

The number of fiber-to-the-home (FTTH) subscribers has rapidly increased in recent years. Triple play services (i.e., fast Internet access, voice over Internet protocol (VoIP), and broadcast video) will be provided on all Internet protocol (IP) networks that are now being standardized as next-generation networks (NXGNs). A NXGN can be realized by extending the IPv6-based Internet to include VoIP and IP multicasting for video distribution. NXGNs are now being deployed, standardized, and financed toward the service starting in 2010. Two major issues for realizing NXGNs are often addressed: how to compromise between transparency and security, and how to provide quality of service (QoS), user authentication, and fairness. As the number of terminals and network traffic increase, however, this IP-based network will face a limit in flexible extendibility owing to the complexity of the network and the huge cost of security assurance.

To solve such problems, research has begun on a new network paradigm, called the new-generation network (NWGN). Network architecture should be studied on the basis of requirements for ubiquitous networking and new networking technologies such as advanced photonic network technologies. A prototype of an NWGN and its standard will be realized in the 2015–2020 time frame. Photonic network technology is expected to serve as a platform for NWGN, with various applications such as real-time video streaming, multipoint video communication, grid computing, digital cinemas, sensor networks, and network games being realized by the seamless control of QoS on a so-called photonic transport platform. The key technologies of the photonic transport platform are data-granularity-adaptive multi-QoS photonic transport, one-hop transparent links, and autonomously controlled power-minimum photonic networks.

QCT should be pursued as one of most promising solutions and be embedded in the photonic transport platform. A practical solution should consist of diverse methods of combining photonic and quantum technologies to realize security at the physical and data layers. This may be referred to as secure photonic network technology. For example, not only restricting us to an OTP via QKD, the secure symmetric key generated by the QKD can be used as the seed key for a contemporary cryptosystem such as an AES, and also for a QNRC to provide reasonable solutions based on quantum technology for high data rate secure communications. Architectures of such quantum secure networks will be a central issue in the next phase of research on quantum cryptography. This is discussed again in Part III.

## I – 5 Initiatives toward Standardization Activities

### 1. Initiatives

The European Telecommunications Standards Institute (ETSI) founded the Industry Specification Group on Quantum Key Distribution and Quantum Technologies (QISG) on July 29th, 2008, and is taking the initiative for the standardization of QKD. This is the result of the SECOQC project under the Sixth Framework Programme of the European Union. The SECOQC consortium immediately selected the ETSI QISG concept as the most suitable and feasible platform for continuation of their work, and the best and easiest option to transfer their results from the FP6-Project SECOQC into standards.

The ETSI QISG is aimed at successfully transferring quantum cryptography out of the controlled and trusted environment of experimental laboratories into the real world where business requirements, malevolent attackers, and social and legal norms have to be respected. The group is currently still in the building phase, today consisting already of some 20 members (global players, small and medium enterprises, research institutes, and universities) based on all five continents.

## 2. Conferences

- Workshop on Quantum Information Technology, London, May 2006 (NIST/Cambridge-MIT Institute's Quantum Technologies Group)
- Discussions at LEOS Summer Topicals, Quebec City, July 2006
- Updating Quantum Cryptography (UQC) 2007, Tokyo, Japan, October 2007 (IPA, NICT, AIST)
- Telcordia One-Day Workshop: Moving toward Requirements for Quantum Key Distribution (QKD), March 3, 2008
- The European Telecommunications Standards Institute, Industry Specification Group on Quantum Key Distribution and Quantum Technologies (QISG) (kicked off on 9 October 2008)
- Updating Quantum Cryptography (UQC) 2008, Tokyo, Japan, December 2008 (IPA, NICT, AIST)

# Part II: Consensual Building of Specifications and Requirements of QCT for Standardization and Commercialization

The current and future eCommunications market can be described as a convergent multimedia market with an increasingly complex structure. Within this market, we are faced with unpredictable, sometimes fragmented, market development (e.g., open network versus a walled garden approach, intelligent networks versus dumb networks) where potential barriers to achieving interoperability may be emerging. Additionally, within the present competitive environment, the risk of noninteroperability is increasing because of, for example, windows of opportunity being small owing to the fast evolution of technology, or the use of non-open standards.

Against this background, there is an ever-increasing awareness on the part of market players and regulators that mass-market development requires interoperability based on open standards. Additionally, the end user appreciates more choice, but expects certainties.

In a world of converging yet diverse technologies, complex systems must communicate and interwork on all levels. This is generally known as interoperability. One well-proven and cost-effective approach to achieve interoperable standards, and subsequently interoperable products, is the holding of interoperability events. These events, which may comprise just a few or many hundreds of participants, draw engineers and equipment into a (possibly distributed) neutral environment where they can execute a large variety of real-life scenarios in various combinations and with different equipment. Successful interoperability events require well-specified tests (scenarios) as well as significant logistical and technical support. However, once in place, such events, or series of events, are an excellent way to validate standards and accelerate standardization. Interoperability events have the additional advantages of optimizing the development of implementations and providing an open forum for resolving issues of noninteroperability and other technical aspects related to the development and validation of standards.

The main aim of standardization is to enable interoperability in a multi-vendor, multi-network, and multi-service environment. Interoperability testing is the structured and formal testing of functions supported remotely by two or more items of equipment communicating by means of standardized protocols. It is not the detailed verification of protocol requirements specified in a conformance test suite, neither is it the less formal development testing often associated with plugfest and interop events (frequently referred to as "bake-offs", see <http://www.interop.com/> for example).

When a new Work Item is raised for a technical specification or a harmonized standard, it is easy to forget the effort that is needed to validate the standard itself to assure the quality of the document and to ensure appropriate test specifications, including those of conformance tests. However, they must be available by the time the standard is implemented as commercial products. This additional effort is rewarded by better standards and the overall improved quality of deliverables.

## II – 1 Security Specifications, Protocols, and Requirements for Secure Communication

In this section, we give an overview of theoretical aspects of secure communications. In section II-1-1, we give basic terminologies that are necessary in this section. Next, in section II-1-2, we list up major QKD protocols and quantum noise randomized cipher as well as the description of the current status of the security of each protocol. In the security analysis, it is important to consider the security not only against any attack, but also against limited attacks since technological difficulties do not allow us to implement any attack. These limited attacks will be described in section II-1-3. In theory, some QKD protocols are shown to achieve unconditionally secure communications, however, practically available devises may not operate in such a way that the theory assumes them to do. These imperfections may be exploited by Eve, and this issue is discussed in section II-1-4. Finally, by summarizing the contents from section II-1-1 to II-1-4, we give a performance specification table in section II-1-5.

### II – 1 – 1 Definitions of Terminologies

**1. Definition: protocol**

The term protocol refers to a description of operations at the abstract level that achieve a cryptographic task. Particularly for QKD protocols, the description usually can be divided into two parts: the first part is quantum communication and the second is classical data processing. The quantum communication part usually specifies

- the relative structure of a set of quantum states that are to be transmitted or exchanged via a quantum channel, and

- the abstract structure of the measurements to be performed by the receiver.

The classical data processing part usually contains

- information that is necessary to distill a secret key from the raw data. It consists of information for sifting, test bits, parity information for error correction, and a hash function for privacy amplification. Information for the sifting highly depends on the quantum communication part; i.e., what kinds of state sets and measurement sets are chosen.

**2. Definition: schemes**

A scheme is defined as a combination of a protocol and the degree of freedom with which the abstract signal states and measurements are to be implemented. For instance, in the case of information encoding in the BB84 protocol, the abstract description of the protocol requires the preparation of four spin-1/2 states: $X$ basis eigenstates $|0x\rangle$ and $|1x\rangle$ and $Y$ basis eigenstates $|0y\rangle$ and $|1y\rangle$. We have several choices for the degree of freedom for the physical implementation; i.e., the encoding scheme. The first is to encode the states into single-photon polarization states, and a second is to use phase encoding. Note that polarization encoding uses two orthogonal polarization modes, so the two basis states are those with one photon in either the first or second mode. The other signal states are then linear superpositions of these states. Similarly, phase encoding uses two orthogonal modes, which correspond to two pulses each, separated by some fixed time. Again, one basis might have a photon either in the first or in the second time window. Typically, one uses the other two bases, where the signal states are equally weighted superpositions of photons in each mode. The basis states then differ only in the relative phase between these two superposition terms.

**3. Definition: security parameters**

Security parameters are defined as parameters describing the degree of the security at an abstract level to be used in information-theoretic security proofs/analysis. The security parameter is chosen by legitimate users, and an example of the security parameter is the number of pulses the sender sends.

**4. Definition: unconditional security**

Roughly speaking, unconditional security means that a protocol is secure against any possible attacks allowed by the law of quantum mechanics. Here, we give two important definitions, the first one is the usual definition of unconditional security and the second one is universal composable security. In this report, whenever we refer to "unconditional security", it means "universal composable security".

- Usual definition of unconditional security

  QKD is unconditionally secure if the following condition is satisfied. For any attack made by Eve, the protocol aborts or succeeds with a high probability $1-O(2^{-s})$, and it is guaranteed that the resulting key is random and Eve's mutual information with the key is less than $O(2^{-l})$. Here, $s$ and $l$ are security parameters that Alice and Bob can choose.

- Universal composable security [Ben-Or et al., 2005; Renner and König, 2005]

  A secret key generated by QKD is used in any cryptographic application, and if we are asked the security of such applications then we have to consider Eve attacking not only the QKD protocol but also its applications. In this scenario, the security analysis would be very difficult as there are many factors. However, it has been shown that if a QKD protocol meets a security condition, which we call the universal composable privacy condition, then the key distilled in the QKD protocol can safely be used for any applications without the degradation of the security of the key.

  For instance, in QKD protocol, Alice and Bob initially share the secret key for the authentication protocol. The universal composable privacy condition guarantees that a secret key generated in a round of the QKD protocol can be securely reused for the next rounds of the authentication protocol in the QKD protocol.

  The usual security criteria, the security proof based on the distillation of a pure state and the one on based on the estimation of Holevo information are known to satisfy the universal composable privacy condition.

### II – 1 – 2 List of Prioritized Schemes and Current Status of Their Security

This subsection is divided mainly into two parts, the first part is devoted to the description of QKD protocols and a technique for increasing achievable secure communication distances. In the second part, we briefly mention quantum noise randomized cipher. In both parts, we review the current status of the security of each protocol.

#### 1. QKD protocols and techniques for increasing achievable secure communication distances

In this subsection, we provide protocols, schemes, and the current status of the security of QKD protocols. We also present the decoy state method, which is a technique for increasing achievable communication distances. The protocol includes quantum communication and classical communication, and as is always the case for any protocol, all the classical communication has to be authenticated. Otherwise, we do not know with whom we perform the QKD protocol. The authentication uses a secret key generated by the previous round of QKD, which means that when we run the QKD protocol for the first time, Alice and Bob must have an initially shared secret key. Thus, the QKD protocol is sometimes called the quantum key growing protocol. We remark that the amount of the secret key needed for the authentication is very small; i.e., $O(LogN)$ for the authentication of $N$ bits [Wegman and Carter, 1981].

## 1A. QKD Protocols

### (1) BB84 protocol

Currently, the most investigated protocol both from the viewpoints of theory and experiment is the BB84 protocol. In what follows, we describe the encoding scheme, signal transmission scheme, measurement scheme, and classical communication for BB84.

#### BB84-A Encoding scheme

As we have already mentioned, experimentally, two encoding schemes have been used so far. The first scheme is polarization encoding and the other is phase encoding of attenuated laser light. In either case, four states (|0x>, |1x>, |0y>, and |1y> or equivalently |H>, |V>, |R>, and |L>) are encoded in the single-photon part.

#### BB84-B Signal transmission scheme

The signal transmission scheme is chosen depending on which encoding scheme is used, otherwise the signal would be subjected to large disturbances. The use of an optical fiber is best in terms of the photon transmission rate; however, birefringence in an optical fiber significantly disturbs the polarization state. Thus, when we use the polarization of a photon (BB84-A1) as the encoding scheme, we usually choose open space as the signal transmission scheme, whereas we use an optical fiber when choosing the relative phase encoding scheme (BB84-A2).

BB84-C Measurement and decoding scheme

The measurement scheme is dependent on the encoding scheme. Again, Bob measures the photon polarization when we use the encoding scheme BB84-A1, and he measures the relative phase when we choose the BB84-A2 encoding scheme. In the latter case, usually a Mach–Zehnder interferometer followed by photon detectors are used.

As part of the measurement, Bob has to randomly choose a measurement basis from two possibilities, the $X$ basis and $Y$ basis. In the case of BB84-A1, the basis is chosen by randomly switching the polarization wave plate, and in the case of BB84-A2, the basis is chosen by randomly applying a phase modulation of either 0 or $\pi/2$ to one arm of the Mach–Zehnder interferometer (see also Fig. 1). In either case, one of the measurement outcomes from each basis, $|0x\rangle$ or $|0y\rangle$ ($|1x\rangle$ or $|1y\rangle$), is interpreted as the detection of the bit value 0 (1).

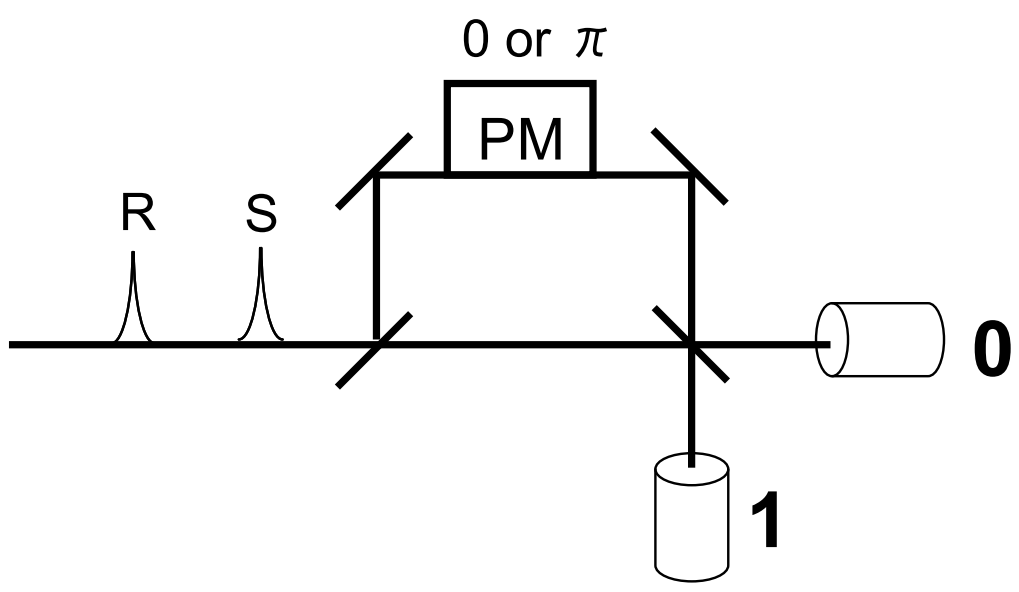

Fig. 1 Decoding scheme using a Mach–Zehnder interferometer. **PM** represents the phase modulator.

BB84-D Classical communication

The classical communication phase is typically divided into four phases: the sifting phase, error/information leakage estimation phase, error reconciliation phase, and privacy amplification phase. Since all the information is exchanged over a public channel and so Eve has free access to it. We note that the sifting phase and details of the leaked information estimation are protocol-dependent, and all other phases are the same among QKD protocols.

BB84-D1 Sifting phase

Alice and Bob exchange the basis they have used for each transmission of the signal, and they keep the events where they have used the same measurement basis and Bob's detector (detectors) clicks (click). At this point, each of them has the "sifted key."

BB84-D2 Error/information leakage estimation phase

To estimate the bit error rate and information leakage, Alice and Bob can broadcast some portions of the sifted key over the public channel. The exposed sifted key is called the test bits, and the test bits give us an idea of the bit error rate, i.e., how much the error rate in the untested bits (code bits) is, and how much information is leaked to Eve. Alternatively, as for the bit error rate, Alice and Bob can perform error correction and learn along the way the actual bit error rate. We remark that the estimation of the leaked information is dependent of what kind of encoding scheme we use as well as the imperfections of the devices, and we do not give a detailed description of the estimation.

BB84-D2a Error reconciliation phase

Based on the estimation of the bit error rate in BB84-D2, Alice and Bob choose and apply an appropriate error correcting/discarding code to make the sifted keys identical.

The error correcting/discarding code is either unidirectional or bidirectional [Assche, 2006, and references therein]. In the unidirectional scheme, only one party sends information to the other party, while both compare and broadcast information in the bidirectional scheme. The information consists of partial information of the code bit, which is used to infer the erroneous bits. Using the information, Alice and Bob can thus correct/discard erroneous bits; however, we note that additional information is leaked to Eve.

Low-Density Parity-Check code is well-known unidirectional code, and the cascade protocol is an example of bidirectional code. After this phase, Alice and Bob share the reconciled key, which has significantly small error. Note that the choice of error correction method (one-way or two-way, direct or reverse reconciliation) has to fit the applied security proof.

BB84-D2b Privacy amplification

To convert the reconciled key into a secure key, Alice and Bob use a privacy amplification protocol. First, using the public channel, they randomly choose a hash function out of a set of hash functions. The set is chosen according to the estimation of the leaked information in BB84-D2. Both Alice and Bob then apply the chosen hash function to the reconciled key so that they share a secret key.

<u>BB84-E Status of the security proof of BB84 and technology required for secure communication</u>

The use of the encoding scheme and the measurement scheme that we described for BB84-A and BB84-C has proven to achieve unconditionally secure communication. We note that when using a single-photon source, the achievable distance is long but the implementation of the single-photon source is technologically difficult. On the other hand, a weak coherent pulse (WCP) source is available using current technology; however, the secure communication distance is compromised. We note that the use of the decoy state method (which we explain later) increases the achievable distance for a WCP, leading to the same key generation scaling as for the single photon.

The detector can be a threshold detector that only discriminates between a vacuum state and non-vacuum state.

### (2) BBM92 protocol [Bennett et al., 1992b]

BBM92 is similar to BB84. One difference is that two legitimate parties perform the Bob's measurement in BB84. These measurements are conducted on the photonic states stemming from a photon-pair source located between Alice and Bob, who can use a threshold detector. Thus, both parties in BBM92 act as Bob does in BB84, following the same measurement and decoding scheme, and all classical information is the same as that in BB84.

The photon-pair source is assumed to be under Eve's control, and Alice and Bob can judge whether they can generate a secret key depending on the bit error rate. In practice, a parametric down-conversion source is often used, where the ideal Bell state in the form $|H\rangle_A|H\rangle_B+|V\rangle_A|V\rangle_B$ is generated with some probability. Here the subscripts $A$ and $B$ represent particles to be distributed to Alice and Bob respectively.

<u>BBM92-E Status of the security proof of BBM92 and technology required for secure communication</u>

The protocol described above is unconditionally secure, with the security proof shown in the same manner as for BB84.

### (3) SARG04 protocol

The experimental implementation of SARG04 is the same as that of BB84. The differences between the two protocols are the definition of the set of the states in the encoding scheme and the definition of the set of the measurements in the decoding scheme as well as details of the estimation of the leaked information. Thus, only the classical communication part is different from what is used in BB84. One of the features that differentiate SARG04 from BB84 is that we can generate a key not only from the single-photon part, but also from the two-photon part. Thus, SARG04 is immune to PNS attack (which we discuss in section II-1-3).

SARG04-A Encoding scheme

As is the case for BB84, two encoding schemes (the polarization encoding and phase encoding schemes) are used for SARG04. Since there is one-to-one correspondence between the two schemes, as an example, we use the polarization scheme to explain SARG04 encoding. In the protocol, Alice sends out the following four photon polarization states.

$$|H\rangle\ ,\ |V\rangle\ ,\ |R\rangle\ ,\ |L\rangle$$

We define four sets of the states. $|H\rangle$ and $|R\rangle$ form "basis **a**", $|R\rangle$ and $|V\rangle$ form "basis **b**", $|V\rangle$ and $|L\rangle$ form "basis **c**", and $|L\rangle$ and $|H\rangle$ form "basis **d**". The first (second) state of each basis encodes the bit value 0 (1). Note that the bit information is encoded in two nonorthogonal states.

SARG04-B Signal transmission scheme

The signal transmission scheme is the same as that outlined for BB84-B.

SARG04-C Measurement and decoding scheme

The experimental measurement scheme is the same as that outlined for BB84-C, and the decoding scheme differs from that of BB84. The information encoded in $|H\rangle$ ($|R\rangle$) of basis **a** is decoded when Bob's measurement outcome is $L$ ($V$). The measurement outcomes $H$ and $R$ are inconclusive. Note that this measurement is an unambiguous state discrimination measurement, and the decoding procedures using other bases are executed in the same manner.

SARG04-D Classical communication

As previously mentioned, the main difference between the classical communications of SARG04 and BB84 is the detail of the estimation of the leaked information, which we do not give the description of.

A minor difference is the sifting phase. In this phase, Alice and Bob exchange bases chosen from four bases, and they keep the events where they have used the same measurement basis and Bob has obtained the conclusive event.

SARG04-E Status of the security proof of SARG04 and technology required for secure communication

The use of the encoding and measurement schemes described in SARG04-A and SARG04-C has been proven to achieve unconditionally secure communication. Moreover, it has been shown that the two-photon emission part can contribute the key generation rate. Note that, in the security proof, we assume the detector discriminates among a vacuum, single photon, and multi-photons. This is a technical requirement in the security proof, which makes actual implementation difficult.

As is the case in BB84, the SARG04 protocol can accommodate the decoy state method to increase the achievable distance of communication.

A comparison between BB84 and SARG04 in terms of the achievable unconditionally secure communication distance shows that BB84 can cover longer distances than SARG04 can, even if we take into account the key generation from the two-photon emission part in the SARG04 protocol.

### (4) B92 protocol

The B92 protocol uses only two nonorthogonal states and is thus the simplest QKD protocol. In principle, any two nonorthogonal states can be used; however, to achieve a longer distance of secure communication, the use of strong reference light is preferable, which we will describe in detail.

B92-A Encoding scheme

The bit value 0 (1) is encoded as two-mode coherent light $|\alpha\rangle_S |\beta\rangle_R$ $(|-\alpha\rangle_S |\beta\rangle_R)$,

and these pulses are chosen randomly. Here, $\alpha$ and $\beta$ can be chosen to be real satisfying $\alpha << \beta$. The signal mode and reference mode are spatially separated, and usually a Mach–Zehnder interferometer together with a phase modulator is used to prepare the pulses.

B92-B Signal transmission scheme

An optical fiber is usually used to efficiently transfer the coherent light.

B92-C Measurement and decoding schemes

Currently, there are two measurement and decoding schemes.

B92-C1 First measurement and decoding scheme

In the first scheme, a Mach–Zehnder interferometer with asymmetric beam splitters is used for the decoding. We use three detectors, two of which are for the decoding and the other for monitoring the reference pulse.

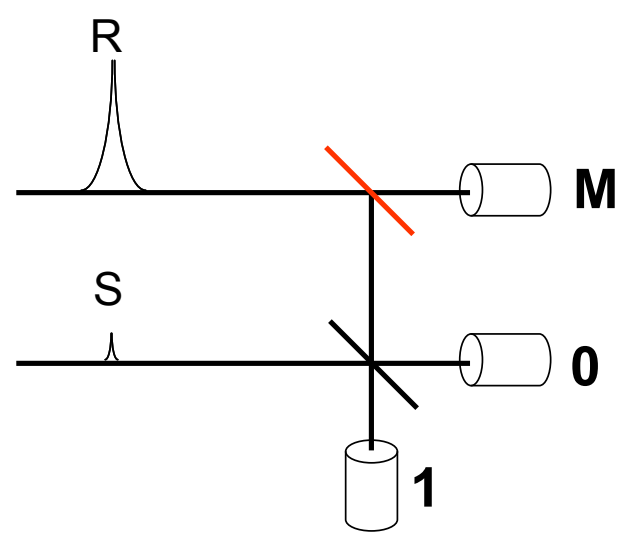

Fig. 2. Schematic diagram of Bob's measurement setup for B92. The red bar represents an asymmetric beam splitter.

Figure 2 is a schematic illustration of Bob's measurement setup where **0** and **1** represent the decoding detectors and **M** the monitoring detector. The beam splitter in front of the monitoring detector is an asymmetric detector, and its transmission rate is tuned in such a way that the mean photon number of the reflected light is the same as that of the incoming signal light in normal operation. The monitoring detector is important for long-distance secure communication since it serves as a countermeasure against unambiguous state discrimination (USD) attack (which we will discuss in section II-1-3). Bob records the bit value 0 (1) when only the detector **0** (**1**) clicks and the monitoring detector **M** detects a photon number in a prefixed photon number regime. He takes note of the ratio of these events as well as other events for the estimation of the leaked information.

B92-C2 Second measurement and decoding scheme

In the second scheme, we assume that Bob has his own local oscillator (strong reference light) and a threshold detector for the bit value reading (see also Fig. 3). First, he measures the relative phase between the incoming strong reference light and his own light, and depending on the measurement outcome, he applies phase modulation to his own reference light. The phase-modulated reference light interferes with the incoming signal light in bit-value decoding. In addition to this feed-forward control, Bob chooses randomly either 0 or $\pi$ as his phase shift, and he applies the phase shift to the signal pulse. This phase shift represents his bit value, and if the threshold detector clicks when he sets the phase shift as 0 ($\pi$), then he regards the outcome as 0 (1). No click is regarded as an inconclusive event. He takes note both of the conclusive event and the ratio of inconclusive events.

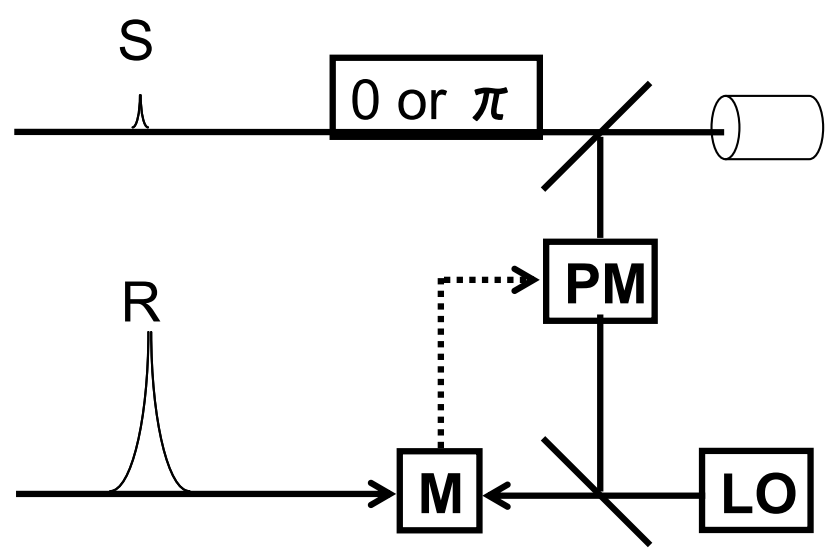

Fig. 3. Measurement setup for the second decoding scheme for B92. **PM** is the phase modulator for the feed-forward control, and the detector is a threshold detector. **LO** represents Bob's local oscillator (reference light).

B92-D Classical communication

The main differences in the classical communication of B92 compared with that of BB84 are the sifting phase and the classical information that is exchanged for the estimation of the leaked information. As is also the case for SARG04, the detail of the estimation of the leaked information differs, which we do not give the description of . All the other classical communication phases run in the same manner as those in BB84 does. Thus, we only mention the sifting phase.

B92-D1 Sifting phase

Bob tells Alice over the public channel which pulse he got conclusive events from. Alice keeps all data for which Bob obtained a conclusive event, and she discards all other data.

B92-E Status of the security proof of B92 and technology required for secure communication

The use of an encoding scheme and the first measurement scheme that we described has proven to achieve unconditionally secure communication. In addition, the use of the second measurement scheme was shown to be unconditionally secure assuming a large limit for the amplitude of the reference light. We note that in the first measurement setup, the detector for the decoding has to discriminate among a vacuum, single photon, and multi-photon, and the monitoring detector can determine whether the detected photon number is inside a particular photon number regime or not. These are technical requirements of the security proof, which makes the actual implementation of the first measurement scheme difficult. On the other hand, as we have mentioned, we can use a threshold detector for the first measurement scheme at the cost of feed-forward control.

A theoretical comparison between BB84 with decoy states and B92 in terms of the achievable unconditionally secure communication distance shows that B92 covers longer distances, which are comparable to those of BB84 with decoy states.

### (5) DPS-QKD

In DPS-QKD, Alice emits trains of pulses, and bit information is encoded into the relative phase between any two consecutive pulses. Thus, a pulse can be both a signal pulse and a reference pulse of the following pulse. Intuitively, this protocol is considered to be strong against the PNS attack presented in section II-1-3 since the time slot in which Bob detects a signal is not under the control of either Bob or Eve.

DPS-A Encoding scheme

Alice emits trains of coherent pulses in such a way that bit information is encoded into the relative phase between two consecutive pulses. A relative phase of 0 ($\pi$) represents a bit value of 0 (1). For instance, the bit string 010 is encoded into a train of four pulses as $|\alpha\rangle_1 |\alpha\rangle_2 |-\alpha\rangle_3 |-\alpha\rangle_4$ or $|-\alpha\rangle_1 |-\alpha\rangle_2 |\alpha\rangle_3 |\alpha\rangle_4$. The mode subscripts represent the time slot of each pulse.

DPS -B Signal transmission scheme

An optical fiber is used to efficiently transfer the coherent light.

DPS -C Measurement and decoding scheme

A Mach–Zehnder interferometer with a delay of one time slot is used to read out the relative phase. We use two detectors that correspond to bit values of 0 and 1.

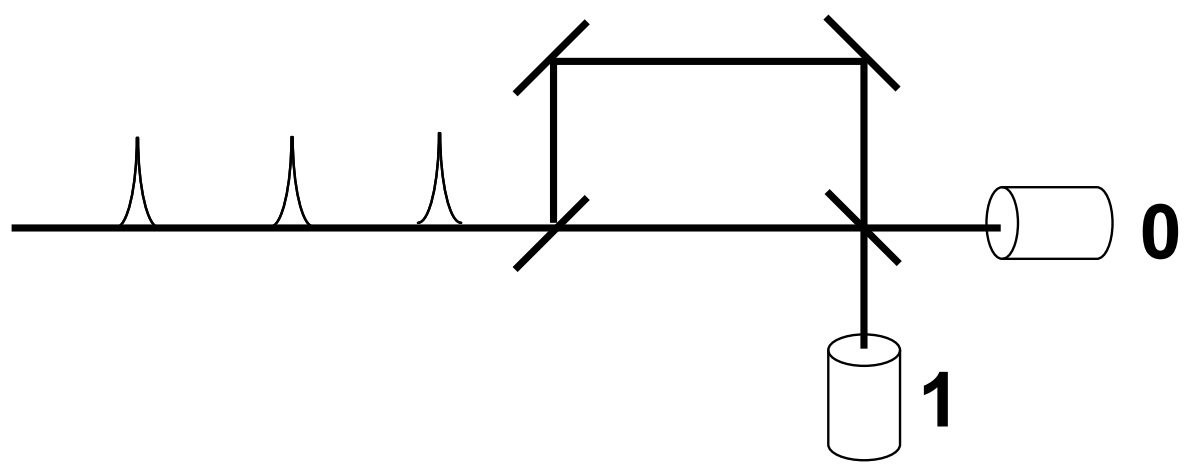

Fig. 4 Schematic diagram of Bob's measurement setup of DPS. The path difference in the Mach–Zehnder interferometer is the same as the distance between two consecutive pulses.

Figure 4 schematically illustrates Bob's measurement setup where **0** and **1** represent the decoding detectors, the clicking of which tell Bob the detected bit value. Bob records which detector clicks as well as the time slot in which the click occurs. Simultaneous clicking of the two detectors and neither detector clicking are regarded as inconclusive events. He also takes note of the ratio of these events.

DPS-D Classical communication

The main differences between the classical communication of DPS and that of BB84 are the sifting phase and the classical information for estimation of the leaked information. The detail of the estimation of the leaked information also differs and we do not describe it here. All other classical communication phases run in the same manner as for those in BB84. Thus, we only mention the sifting phase in what follows.

DPS-D1 Sifting phase

Bob tells Alice over the public channel in which time slot he received the conclusive events. Alice keeps all data corresponding to Bob's conclusive events and discards all other data.

DPS-E Status of the security proof of DPS and technology required for secure communication

The unconditional security of DPS has not yet been proven. Instead, security against a general individual attack on a photon in a train of signals has been shown [Waks et al., 2006]. On the other hand, the unconditional security of a DPS scheme with a single-photon source has been proven [Wen et al., 2006]. In this case, the detector has to be one that can discriminate among a vacuum, single photon, and multi-photons.

A sequential attack is known to be effective against a DPS scheme. In such an attack, Eve uses the USD measurement for each pulse, and when she has consecutive successful USD measurement outcomes, she resends signals to Bob [Curty et al., 2007; Tsurumaru, 2007]. These signals are not necessarily coherent light pulses and they can be a superposition state of a single photon. Analysis shows that this attack is effective in the long distance regime.

### (6) CV-QKD

The main difference between the CV-QKD protocol and the above QKD protocols is the use of homodyne-based measurements instead of photon-counting-based measurements for CV-QKD, where the former measures the quadrature amplitude of the signal, which is a continuous variable. Several CV-QKD protocols have been proposed so far on the basis of squeezed states or coherent states as the signals. In what follows, details of CV-QKD protocols based on a coherent state are described.

<u>CV-A Encoding scheme</u>

Random values are encoded in the complex amplitude of the coherent state signal. There are two encoding schemes depending on the modulation format.

CV-A1 Gaussian modulation

Signals are continuously encoded in a complex amplitude α (i.e., two-dimensional phase space) of the coherent state $|\alpha\rangle$. When Bob's measurement is a homodyne measurement, Alice encodes a single-bit value, while she can encode a two-bit value if Bob performs a heterodyne measurement. The latter encoding scheme is referred to as a doubly encoded scheme.

CV-A2 Discrete modulation

The signal is phase modulated by a fixed amount depending on the randomly chosen basis and bit value. For instance, a q-squeezed state is displaced in $q$ by $+c$ ($-c$) when Alice's bit value is 0 (1), where $c$ is a positive constant. The encoding with respect to $p$ goes in the same manner.

CV-B Signal transmission scheme

Coherent states are efficiently transferred via an optical fiber or free-space propagation.

CV-C Measurement and decoding scheme

In the following, we describe the measurement scheme. The decoding scheme is dependent on whether the protocol is discrete-modulated or Gaussian-modulated. Bob tries to readout the discrete variable encoded by Alice in the case of discrete-modulated while he uses a prefixed decoding scheme in the case of Gaussian modulation. An example of the prefixed decoding scheme is as follows. Imagine that Bob measures the $x$ component of the complex amplitude, which is a continuous variable. In this case, Bob interprets the measurement outcome $x$ as the bit value 0 (1) if $n$ is even (odd) where $(2n - 1)c < x \leq (2n + 1)c$ and $n = \ldots -3, -2, -1, 0, 1, 2, 3, \ldots$ Decoding for the doubly encoding scheme is in the same manner with respect to both $x$ and $p$.

CV-C1 Homodyne detection

In this measurement setup, either the $x$ or $p$ component of the complex amplitude is randomly read out. This random choice of the basis corresponds to the random basis choice in BB84.

CV-C2 Heterodyne detection

By splitting the incoming light, both the $x$ and $p$ components of the complex amplitude are randomly read out. Note that this does not mean both $x$ and $p$ components can be measured simultaneously. The splitting, of course, enlarges the variance of the statistic of the measurement outcomes.

CV-D Classical communication

CV-D1 Sifting phase

If the protocol involves a random choice of the basis, then the sifting process must be performed in the same manner as for the sifting phase in BB84. However, if the protocol is a doubly encoding scheme, then we do not need to perform the sifting.

CV-D2 Noise/information leakage estimation phase

To estimate channel noise and information leakage, Alice and Bob sacrifice portions of the sifted key as test bits, and Bob broadcasts all the relevant statistics he took note of in the measurement stage. We do not give the detailed description of the estimation of the leakage information.

CV-D2a Error reconciliation phase

In the one-way error correction, the syndrome goes from Alice to Bob (direct reconciliation) or from Bob to Alice (reverse reconciliation).

<u>CV-E Status of the security proof of CV-QKD and technology required for secure communication</u>

There are many versions of CV protocols, and some of them are unconditionally secure.

**1B Technique for increasing the achievable distance of secure communication (decoy state method)**

In this subsection, we describe the decoy state method, which increases the achievable distance of secure communication. This method is used for single-photon-based QKD, such as BB84, SARG04, six-state protocol (the six-state version of BB84), and the single-photon-based B92 with imperfect source implementation to combat the PNS attack that will be discussed in section II-1-3. This method increases secure communication distances when the light source is one that sometimes emits multi-photon, leading to the same scaling of the key generation rate as for the perfect single-photon source.

In this method, the encoding scheme, signal transmission scheme, measurement scheme, and decoding scheme are the same as the original protocols; i.e., BB84 or SARG04. However, there are some differences. The first difference is that Alice randomly chooses and emits decoy pulses that have mean photon number that is different from the one of the signal pulse. Typically, the number of choices for the decoy pulses is two, three, or four including the vacuum state. The basic idea behind this method is that Eve cannot discriminate whether the received pulse is a decoy pulse or not. Thus, she cannot behave differently for a signal pulse and decoy pulse, and what Eve does to the decoy pulses gives us a clue about what she has done to the single-photon part of the signal pulses. More precisely, Alice and Bob expose all the data stemmed from the decoy pulses to estimate the bit error rate and the ratio of the detection events for each decoy pulse. This data processing for the estimation serves as the second difference, and this estimation gives us an idea of the bit error rate and the ratio of the detection events for the single-photon part of the signal pulse, which is taken into account in the privacy amplification.

## 2. QNRC

Besides QKD schemes, the QNRC protocol also uses quantum properties to make communication substantially secure. We define the class of the protocol as one-way secret communication protocols with physical encoding schemes that directly encode messages into a particular quantum state depending on the message, secret key information, and other physical inner states of the encoder. The use of quantum noise would reinforce secrecy in communication and increase the efficiency in the use of the communication bandwidth, although cryptographic understanding of the reinforced secrecy is an issue to be carefully investigated. Here we describe schemes and the current status of the security of a QNRC based on the Y00 (or αη) protocol, which is a protocol known to realize the above concept.

Y00-A Key expansion scheme

Alice and Bob share a secret key (referred to as the seed key). The seed key is expanded by a key expansion function (e.g., a linear feedback shift register or AES in stream cipher mode) to generate a running key sequence ($Z_1$, $Z_2$ ...) such that each $Z_i$ takes $M$ values.

Y00-B Encryption scheme

Length $n$ plaintext ($X_1$, $X_2$ ...) is encoded in the coherent state:

$$|\psi(X_i, Z_i)\rangle = |\alpha \exp\{i\theta(X_i, Z_i)\}\rangle$$

$$\theta(X_i, Z_i) = [Z_i / M + (X_i \oplus Pol(Z_i))]\pi$$

where $Pol(Z_i)=0$ or 1 according to whether $Z_i$ is even or odd. Note that one can also encrypt text in coherent states with different physical formats such as DPS keying or intensity modulation.

Y00-C Signal transmission scheme

Coherent states are efficiently transferred via an optical fiber or free-space propagation.

Y00-D Decryption scheme

Bob generates a running key sequence ($Z_1$, $Z_2$ ...) from the seed key and the same key expansion function as Alice uses. Then for each quantum signal $|\psi(X_i, Z_i)\rangle$, Bob makes a homodyne measurement to discriminate $|\psi(0, Z_i)\rangle$ and $|\psi(1, Z_i)\rangle$ and decrypt $X_i$.

Y00-E Status of the security of Y00

In practice, the security of Y00 and related protocols is based on computational complexity. As mentioned in section I-2-1, quantitative security analyses of the protocols have not yet produced clear results. The protocols are at least as secure as conventional (nonrandom) ciphers, but how much the additional randomness due to quantum noise increases the complexity is still unknown. It is highly desirable to conduct contemporary cipher analyses.

### II – 1 – 3 Limited Attacks

It is important to consider the unconditional security, however, because of the technological limitation, we cannot actually implement any attack. Therefore, for the practical purpose, it is important to consider the security against some limited attacks in order to see the secure communication distances that a protocol can achieve from the practical viewpoint. The analysis of a particular attack sometimes reveals the essential fragility of a protocol, and this fragility serves as a hint to modify the protocol or to consider the countermeasure against the attack. Thus, such analysis not only gives us the limitation of a protocol, but it is also helpful to improve the protocol. In this subsection, we list up several limited attacks against QKD and QNRC.

#### 1. QKD

Here we introduce classes of attacks against QKD, followed by limited attacks.

Classes of attacks

- Collective Attacks [Biham and Mor, 1997]
  In collective attacks, Eve first prepares ancilla systems, each of which is to interact with each signal sent by Alice. After the interactions and listening to classical information over the public channel, Eve performs an optimal joint measurement on all ancilla systems to retrieve information on the key. This attack class has been introduced since it eases theoretic analysis. We now know that security against this type of attack also implies security against most general coherent attacks [Renner and Cirac, 2008].

- Individual Attacks [Lütkenhaus, 2000]

  In individual attacks, Eve first prepares ancilla systems, each of which is to interact with each signal sent by Alice. After the interactions and listening to classical information over the public channel, Eve performs an optimal individual measurement on each ancilla system. Thus, the difference between the collective attacks and the individual attacks is whether Eve's final measurement is a collective measurement or not. The operational definition would be that Eve does not require interacting quantum memories for this attack, though a variation of individual attacks without a delay of the measurement might be by now a more reasonable choice.

## Limited attacks

- Photon Number Splitting (PNS) Attack [Bennett et al., 1992; Huttner et al., 1995]

  The PNS attack was invented to eavesdrop on the BB84 protocol or six-state protocol. Original proposals of these protocols assumed the use of a single-photon source; however, since the implementation of the single-photon source is difficult, the attenuated laser light source is replaced with it. The crucial difference between a single-photon source and laser light source is that multi-photon is emitted by the laser light source. From the multi-photon emission part, Eve can obtain information without causing disturbance using the following attack, which we refer to as the PNS attack.

  First, Eve performs a quantum nondemolition measurement of the photon number, and if the number is greater than one, say $n$ (>1), then she keeps $n-1$ photons while letting the single photon go to Bob's side. After listening to the basis information exchanged over the public channel, Eve conducts the measurement with the basis so that she can obtain bit information without causing errors.

  In the security analysis, we make the worst case assumption that Eve sends as many signals as possible to Bob from Alice's multi-photon emission part, while she suppresses the sending of signals from Alice's single-photon emission part as much as possible. With this assumption, the resulting secure communication distance turns out to be very short. Thus, the use of an attenuated laser light source limits the secure communication distance.

This attack is an important example of the vulnerability of imperfect devices. We note that this attack is applicable for any photon source that emits multi-photon. Finally, we note that the decoy state method is a good countermeasure of this attack, and it allows us to increase the secure communication distance as we do not need to work on the worst case assumption in the security analysis.

- Unambiguous State Discrimination (USD) Attack [Dûsek et al., 2000]

  A natural way to obtain information from the incoming signals sent by Alice is to try to identify the state. It is known that if the states are linearly independent of one another, then state identification is possible with some probability. This type of measurement is called USD, which is one of the most important tools for eavesdropping. In principle, Eve can utilize this measurement for eavesdropping on the B92, DPS-QKD, and BB84 protocols without phase randomization where linearly independent states are used. It is known that a USD attack significantly limits the secure communication distance of BB84 without phase randomization compared with BB84 with phase randomization; however, the limitation that USD poses on DPS-QKD is not so significant. Regarding B92, if this protocol is implemented with a strong reference light, then USD attack is not effective because the failure identification in USD results in bit errors. Typically, USD attacks show the limit of the performance of schemes in the absence of errors.

- Intercept and Resend Attack

  Among attacks described in this subsection, the intercept and resend attack is the most practical. In this attack, Eve makes a measurement, and depending on the measurement outcome, she resends states to Bob. The measurement might be the same as the one Bob performs and the states to be resent are those Alice prepares. In this case, Eve impersonates both Alice and Bob, which is obviously feasible with current technologies. Since Eve's operation is equivalent to the so-called entanglement breaking channel, this attack does not allow the distribution of quantum correlations needed to establish a secret key [Curty et al., 2004]. Thus, QKD protocol that can be broken by an intercept and resend attack cannot be considered secure no matter what modification Alice and Bob make in the classical communication part.

**2. QNRC**

As already mentioned in previous sections, the QNRC is expected to provide higher computational complex-theoretic security than contemporary symmetric ciphers do. Therefore, the security notions listed here follow those commonly used for contemporary symmetric ciphers. Typical security analyses of symmetric ciphers are classified into two categories depending on the scenario [Katz and Lindell, 2007].

- Ciphertext only attack
  For the attacker, whose goal is to obtain information of both the message and secret key, only ciphertext is available in the first scenario. In the case of quantum data encoding schemes where the definition of ciphertext might be ambiguous, we suppose that the quantum state of the physical carrier is available to the attacker. When the encoded message is random, the information of the secret key must not be leaked in an informational sense whatever the security of the scheme.

- Known plaintext attack
  We suppose the information of the message as well as the corresponding ciphertext is available to the attacker. The goal of the attacker is to obtain information of the secret key.

### II – 1 – 4 Security Threats and Imperfections of Devices

In this subsection, we briefly discuss security threats/imperfections of devices and problems that need to be resolved for the realization of secure and fast communications. In a security proof, we usually assume that Bob's two detectors for the bit value reading have the same quantum efficiencies. However, this is difficult to accomplish, and normally two detectors have different quantum efficiencies, which Eve may exploit for eavesdropping. Another example of a problem arising from a device imperfection is a side channel. Imagine that Alice performs phase modulations on the signal pulses depending on the chosen bit value. In this case, if the phase modulation emits some electromagnetic wave dependent on the bit value, then Eve can simply measure the electromagnetic wave to obtain bit information and she does not need to attack the QKD itself. The side-channel effects exist regardless of quantum and classical cryptography, and thus they are very difficult to eliminate. In the following, we list possible threats and problems that need to be taken into account for secure and fast communications.

*State Preparation*

- Source evaluations: Most security proofs assume the characterization of the source. Thus, we need to consider how to experimentally check or characterize the state of outgoing signal pulses.
- Random number generator (RNG): A fast and true random number generator is necessary for the state preparation in most of QKD protocols. Such a generator is very difficult to realize.
- Strong pulse attack: Eve might shed strong light onto Alice's side to peep at which phase modulation Alice applies. By measuring the reflected light, Eve can infer the encoded information. We need to consider a theoretical/experimental countermeasure against this attack.
- Side-channel vulnerability: There might be a case in which unwanted signals containing useful information, such as bit information and basis information, are emitted from the sender's devices. Theoretical and experimental countermeasures against this attack need to be considered.

*Measurement*

- Measurement unit characterization: Most security proofs assume a mathematical description of the measurement. However, given the experimental devices, it is very difficult to write a precise mathematical representation.
- Imperfections of devices: In most cases, assumptions in theory are not satisfied in experiments. One example is a detector efficiency mismatch. Such imperfections must be taken into account theoretically and/or removed experimentally.
- Side-channel vulnerability: This is the same as side-channel vulnerability in the state preparation.
- RNG: This is the same as the issue for the RNG in the state preparation.

*Classical Communication*

- Finite size effect: Most theory of security analyses assumes that the number of pulses Alice emits is infinite, which makes the analysis simpler as we do not need to consider variances of data. In practice however, the number of pluses is finite, and the finite size effect has to be treated seriously.
- Error correction: A classical error correcting code that is efficient in terms of the decoding speed and the capability of correcting as many errors as possible is needed for fast communication or long-distance communication.

- Privacy amplification: This is related to the RNG problem. In privacy amplification for fast communication, we need to generate a random number and make the calculations very fast.

### II – 1 – 5 Performance Specification Table

Following the discussion from section II-1-1 to II-1-4, the assumptions and performances of QKD protocols may be summarized in the following table.

| Protocol | | | Assumptions | | | Performances | | | | | Remark |
|---|---|---|---|---|---|---|---|---|---|---|---|
| | | source | threshold detector | detector effciency | source state | secure against (ideal case) | raw bit rate | $R \propto \alpha^{-x}$ | short distance | long haul | |
| entanglement based (EB) | E91 | EPR | OK | | self-tested? | coherent | Low(Source pair creation) | 1 | ○ | ○ | violation of Bell inequality; fidelity to EPR |
| | BBM92 | EPR | | identical | imperfect | coherent | L (S. pair creation) | 1 | ○ | ○ | same as BB84 |
| prepare & measurement (PM) | BB84 | single photo | OK | can be differer | imperfect | coherent | L (S. emission) | 1 | ○ | ○ | |
| | | WCP | OK | can be differer | imperfect | coherent | Med (Det. afterpulse) | 2 | ○ | × | |
| | | WCP (decoy | OK | can be differer | imperfect | coherent | M (Det. afterpulse) | 1 | △ | ○ | |
| | six-state | single photo | No Squash Op. | identical | perfect | coherent | L (S. emission) (2/3) of BB84 | 1 | ○ | ○ | |
| | | WCP | No Squash | identical | perfect | coherent | M (Det. afterpulse) | 2 | △ | ○ (decoy) | |
| | B92 | WCP (w ref. | OK(Ko)/ NG(Ta) | identical | perfect | coherent | L (inconc) | 1 | △ | ○ | single photon B92 covers only short |
| | SARG | WCP | | identical | perfect | coherent | M (Det. afterpulse) <BB84 | 1.5 | △ | ○ (decoy) | |
| | DPS | single photon/ WCP | | identical | perfect | individual(**) | M (Det. afterpulse) | ? (1: single-photon | ○ | ○ | simple implementation |
| | COW | WCP | | identical | perfect | individual | M (Det. afterpulse) | ? | ○ | ○ | |
| | CV | | homodyne | identical | perfect | collective(*) | High? (Det. S/N) | ? | ○ | × | heavy key |

* coherent for small channel loss ○:applicable, △:ineffective,
** coherent for single photon DPS ×:not recommended

## II – 2 Interoperability Specifications and Requirements

### II – 2 – 1 Interoperability with a Contemporary Cryptographic System

There are two possible problems with the interoperability of QKD and a contemporary cryptographic system, and they may be topics in future UQC meetings. The first problem concerns the loss of unconditional security that occurs when one combines QKD with modern cryptographic algorithms, such as an AES, for encrypting messages. This setup is commonly used in most of today's QKD products or in experimental demonstrations (e.g., for updating an AES key every 5 minutes) mainly because key generations of a QKD are not yet fast enough in some applications to catch up with communication using OTPs. Although unconditional security is no longer guaranteed in these cases, it may still be possible that refreshing secret keys at short intervals can enhance the security of conventional cryptographic systems. However, secret keys are not updated in almost all secret communication using conventional cryptographic systems. As far as we know, there are no theories that justify such intuition, which might be done by introducing new security criteria.

The second problem is the combination of QKD and information-theoretic secure cryptographic protocols. In the field of modern cryptography, once a large number of secret keys are available, there are many protocols that realize cryptographic functions other than secret communications while achieving information-theoretic security. There are important problems in the application of a protocol that provides mutual authentication or a digital signature within a conventional cryptographic system from the viewpoint of an application program interface (API) and the data format. In addition, the concept and security evaluation methods in conventional cryptographic protocols differ from those for QKD and information-theoretic secure cryptographic protocols. We need to determine the achieved security function of such combination systems and security requirements.

### II – 2 – 2 Interoperability among Quantum Cryptosystems

**1. QKD networks: general considerations**

Different schemes of quantum cryptography need to be integrated so as to provide a wide range of solutions for highly secure networks, rather than users being restricted to an OTP by QKD. For example, a secure symmetric key generated by QKD can be used as the seed key for quantum noise-randomized encryption to provide reasonable solutions based on quantum technology for high data rate secure communications. Architectures of such quantum secure networks will be a central issue in the next phase of research on quantum cryptography. In this subsection, we focus on the first necessary step to realize QKD networks having as long a range as possible.

QKD links can only operate over point-to-point connections between two users, and cannot be deployed over any arbitrary network topology. To overcome this limitation, it is important to realize networking QKD links or QKD networks between multiple users. Interoperability among different quantum cryptosystems is particularly indispensable for the effective integration of QKD into secure networks.

For sound cryptographic and security analysis as well to focus on the essential goals of a QKD network, it is important to determine the main properties and objectives of such a network. In general, there are different ways to define a QKD network. One QKD network concept is of an infrastructure for information-theoretic secure key agreement, which relies on quantum resources available to the legitimate participants, while not imposing bounds on the eavesdropping capabilities of the adversary, and allows the connectivity of parties that do not share a direct, fixed quantum channel. This definition naturally extends the properties of a point-to-point (link) QKD. A QKD network specified as above could in principle allow (depending on the realization) the lifting of the typical restrictions for stand-alone QKD links and establish key sharing over long distances (e.g., on a continental scale) by increasing and maximizing the throughput capacity (the key generation rate), ensuring robustness against denial of service attacks and technical service breakdowns.

Two techniques, quantum channel switching and trusted repetition, are available for constructing a QKD network. An end-to-end quantum channel over many nodes can in principle be created using quantum repeaters that have not yet been technically realized. Current technology allows for optical switching and/or trusted repeater networks.

**(a) Quantum channel switching**

Quantum channel switching can create an end-to-end quantum channel (or more generally distributed quantum resources) between Alice and Bob. In optically switched quantum networks, some classical optical functions such as beam splitting, switching, multiplexing, and demultiplexing can be applied to the quantum signals to create a direct quantum channel physically (i.e., on demand). The interest in such optical networking capabilities in the context of QKD networks is that they allow us to go beyond two-user QKD. Active optical switching can be used to allow the selective connection of any two parties with a direct quantum channel. Optical functions can thus be used to realize multiple-user QKD, and the intermediate sites do not need to be trusted since quantum signals are transmitted over a quantum channel with no interruption from one end-user QKD device to another. In this sense, the security analysis coincides with that for a stand-alone QKD link.

This QKD network model cannot, however, be used to extend the distance over which keys can be distributed. Indeed, the extra optical losses introduced in the switching devices will in reality decrease the transmission capacity of quantum channels and thus the maximal key distribution distance. In addition, in a fully switched optical network, any two parties need to share an initial secret so as to be able to start the key agreement process. Overall, these types of networks are not scalable and thus not suitable for a long distance QKD network.

From the point of view of interoperability, such networks require QKD devices of exactly the same type, and in fact, from the same vendor. For example, one "Alice" device can be connected over a passive switch (a beam splitter) to two "Bob" devices of the same type. The set of quantum signals would be split into two subsets, Bob1 and Bob2, allowing Alice, after sifting, to establish two independent keys with the two Bobs. Active switching would in turn allow a direct communication (like a classical telephone line) between an "Alice" device and a "Bob" device. In summary, such solutions would allow any-to-any communication on a metropolitan scale.

**(b) Trusted repeater**

A trusted repeater transports keys over many intermediate nodes, which are trustworthy (i.e., not infiltrated by an eavesdropper). Trusted repeater QKD networks have been discussed in various contexts since the advent of quantum cryptography. Essentially, these networks are infrastructures composed of QKD links (pairs of QKD devices associated by a quantum and a classical communication channel), each link connecting two separate locations (nodes). A QKD trusted repeater network is then a connected graph, the vertices of which are nodes, and the edges QKD links. Several QKD devices—the end points of links pointing to different nodes—are then accumulated in a single node. Using the links connected to a node, it is possible to retransmit (repeat) secret information along the network. The goal is network-wide key distribution whereby the distributed key does not necessarily have a QKD origin. To avoid confusion, this key is referred to as "secret" in what follows. A particular mechanism (sometimes called a hop-by-hop mechanism) works as follows.

- The nodes are equipped with classical memories for accumulating the key material generated over the connected QKD links.
- Secret distribution is performed over a QKD path (i.e., a one-dimensional chain of QKD links and corresponding nodes), establishing a connection between a sender node and recipient node.
- Secrets (e.g., those generated by a true RNG) are forwarded using unconditionally secure transport along the path. At each node, the outgoing secret is encrypted by an OTP using key material which was previously generated over the outgoing QKD link from the chain and stored in the memory of the node.

- The resulting cipher message is classically dispatched together with an authentication tag to the next node on the path; i.e., to the one connected to the other end of the very same link. After the received authentication tag is verified, the transported secret is then decrypted using the same key material as that used for encryption. This material is stored in the memory of the node and naturally originates from the incoming QKD link. The process is repeated until the transported secret reaches its destination.

End-to-end information-theoretic security is obtained between the sender and recipient nodes provided that all intermediate nodes can be trusted, as the nodes possess the full communicated information. The trusted nodes thus play the role of (classical) trusted repeaters.

Generally speaking, trusted repeater QKD networks allow the covering of arbitrary distances, connection of an arbitrary number of participants, and use of pairs of QKD devices of different types and from different vendors.

The first proof-of-principle QKD network demonstrator, the DARPA Quantum Network, was deployed between Harvard University, Boston University, and BBN Technologies in 2004 [Elliott, 2002; Elliott et al., 2005] involving both switching and trusted repeater techniques (see below). A highly integrated trusted repeater network demonstrator, developed within the framework of the Integrated European Project SECOQC (financed by the European Commission within the framework of FP6 between 2004 and 2008) was deployed, tested, and demonstrated in Vienna [Poppe et al., 2008].

**2. Example of QKD networks: SECOQC trusted network demonstrator**

The SECOQC trusted repeater network demonstrator (Vienna, October 2008) consists of eight different QKD links of six different types: three Plug-and-Play systems from idQuantique, a Coherent One-Way system from GAP Optique with the participation of idQuantique and the Austrian Research Centers, a CV system from Centre National de la Recherche Scientifique and THALES Research and Technology with the participation of Universite Libre des Brussels, a One-Way Weak Pulse System from Toshiba Research of the United Kingdom, and an Entangled Photons System from the University of Vienna and the Austrian Research Centers. Additionally, two nodes situated in adjacent buildings were connected by a free-space link from Ludwig Maximillians University in Munich (line of sight of 81 m). The QKD links were integrated into a network consisting of six nodes. The average distance between the nodes is between 20 and 30 kilometers, with the longest link being 83 kilometers.

SECOQC has introduced the additional constraint that initial secret keys

(needed for authentication) are only shared between neighboring nodes (i.e., nodes directly connected by a QKD link) and not between any arbitrary pair. This constraint ensures that the number of initial secrets to be shared scales (for wide-area networks) with the number of network nodes and not with their square. This in turn largely simplifies the initialization of a QKD network and the adoption of additional nodes during operation.

An essential feature of the prototype is its network architecture. The corner stone of this architecture is the design of the node. As outlined above, the node essentially contains entities that manage the keys generated over QKD links and ensures cryptographic services (encryption and authentication) for the transport of secret information. At the same time, each QKD device is equipped with a mechanism for device-to-device classical communication, key management (initial and subsequent authentication keys), and cryptographic services (authentication) so as to have the capacity of distilling a key. To overcome this redundancy, SECOQC has put forward the following approach. QKD devices are stripped of stand-alone functionality. They have access to the quantum channel alone and classical communication with a dedicated node device, called a node module (designed and implemented by the Austrian Research Centers). The node module manages all key material of the underlying QKD device and provides an authenticated classical channel for the device. In this sense, the only objective of the QKD device is to communicate over the quantum channel and distill and push a key to the node, using the communication facilities of the latter. The node in turn manages the point-to-point connections (including classical communication with neighbors, key management, and cryptographic services) to be in a position to find paths to the required destinations and realize secure transport protocols as outlined above.

Three types of services can be grouped in network layers: a quantum point-to-point (Q3P) layer, a quantum network layer, and a quantum transport layer [Dianati et al., 2008]. The Q3P layer encapsulates the specifics of QKD. For this layer, QKD devices are the key providers. The upper layers use the Q3P layer alone and have no information on the underlying technology. This design allows a seamless integration of arbitrary QKD devices into QKD trusted repeater networks. Technically, SECOQC has put forward a standardized Q3P specification and defined interfaces for device-to-node-module interoperation. As already mentioned, the cost is the loss of stand-alone functionality of the SECOQC QKD devices. However, such functionality is easily regained by adding two node modules (running only Q3P layers) to such devices or by simply integrating Q3P software into each device.

The quantum network and quantum transport layers are generally similar to those of existing classical network protocols. However, two issues are of special importance. To ensure network-wide information-theoretic security, key material transport has to be information-theoretic secure. Simultaneous path finding and transport protocols require considerable auxiliary network traffic (signaling). It would be a significant waste of key material to use information-theoretic secure transport for this entire load. A nontrivial cryptographic task is to determine optimal approaches and potential attacks on this level. In SECOQC, the respective protocols use authentic transmission of the signaling information and information-theoretic secure transmission for the payload.

It should be emphasized that all three network layers have been designed so that the only target of communication is end-to-end key distribution. Secure communication of the end-user information is handed over to a classic (or generally, any type of) secure communication infrastructure. The latter could use at will the key distributed between the secure locations of the QKD network.

This approach effectively defines three separate network planes: a quantum plane (quantum channels and QKD devices that push the key to the node modules), a secret information plane (node modules with classical communication channels between them with Q3P, network, and transport logical layers that use the QKD-generated key to distribute an information-theoretic secure key between any two nodes on the network), and a data plane (in which the distributed key is used by a secure communication infrastructure to ensure end-to-end network secure communication).

While the SECOQC approach allows complete interoperability in the trusted repeater regime, a development of adequate QKD network architecture in the mixed regime (e.g., trusted repeater backbones combined with local area switched networks) remains an open issue requiring further dedicated research.

### II – 3 Derived Test Requirements

We here consider testing and measurement (T&M) procedures for a QKD system. There are three aspects to T&M: ensuring security by confirming the assumptions behind the security proof, increasing the final key rate by restricting the eavesdropper’s knowledge, and achieving stable operation. The first and second aspects are closely related; security proofs with a small number of assumptions may overestimate the leakage of information to Eve and yield only a low final key rate. We treat the first two aspects together, and discuss separately the third aspect on calibration and synchronization. Subtle issues arise because T&M procedures should never give clues to the eavesdropper. To reduce the burden of a secure T&M procedure, it is necessary to select a protocol, design robust implementation, and develop stable devices.

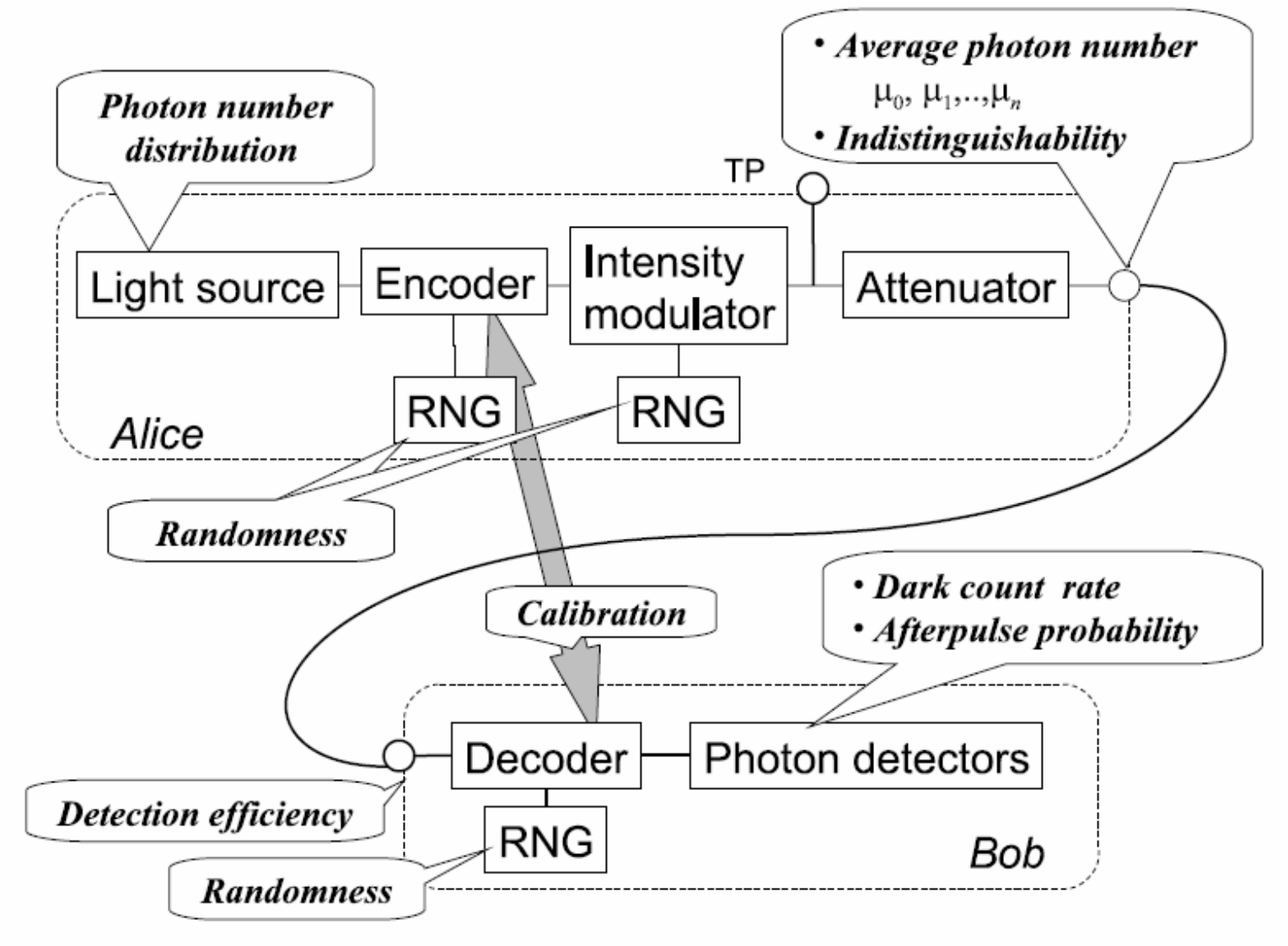

**Figure 1.** A reference model of a QKD system implementing decoy-BB84 protocol. Test items are shown in the balloons.

## II – 3 – 1 Reference System

The following consideration assumes the use of the decoy-BB84 protocol [Hwang, 2003; Lo et al., 2005b; Wang, 2005], which has been extensively studied for practical QKD systems. Knowledge gained can be applied to systems based on other protocols. A reference system is helpful in considering the effects of imperfections in practical QKD systems. We here define a reference system that includes device imperfections in a tractable form. The system is depicted in Fig. 1 and consists of a transmitter, quantum channel, and receiver.

In the transmitter, a light source, which we assume to be a laser, emits light pulses with a Poissonian photon number distribution and an average photon number $\mu_L = |\alpha_L|^2$. There should be no phase relation among the successive pulses (i.e., no phase reference). An encoder then creates four photon states for the BB84 protocol according to the output of an RNG. In $X$–$Y$ coding, the photon states after the modulator are represented by the points (100), (–100), (010), and (0–10) on the Bloch sphere:

$$|\Psi\rangle_{XY} = |\alpha_L/\sqrt{2}\rangle_F |\alpha_L \exp[i\phi]/\sqrt{2}\rangle_S \qquad \text{(II-3.1)}$$

where the relative phase $\phi$ between the fast[1] component $F$ and the slow component $S$

[1] We assume a time-divided double pulse, which is common in fiber-based QKD systems.

varies over $\phi = 0, \pi, \pi/2$, and $3\pi/2$. In $X$–$Y$ coding, the photon states are represented by the points (100), (–100), (001), and (00–1) on the Bloch sphere:

$$|\Psi\rangle_{XZ} = |\alpha_F\rangle_F |\alpha_S \exp[i\phi]\rangle_S \qquad \text{(II-3.2)}$$

where $(\alpha_F, \alpha_S) = (\alpha_L, 0), (0, \alpha_L)$ for $Z$ states and $\alpha_F = \alpha_S = \alpha_L/\sqrt{2}$, $\phi = 0, \pi$ for $X$ states. The average photon number of the pulses should be randomly selected from $\mu = 0, \mu_1, \ldots, \mu_n$ by an intensity modulator. Intensity modulators should change the attenuation for each pulse so that the bandwidth exceeds the clock frequency.

In the receiver, a decoder analyzes the input states and the photon is transported to one of the photon detectors according to its state. The analyzing basis of the input state may be selected by a random number (active choice) or by a beam splitter (passive choice). We assume the photon detectors to be threshold devices (not photon-number resolving), whose detection events originate from photon detection with finite detection efficiency and dark counts.

### II – 3 – 2 Test Items for Secure Key Generation

The T&M items are shown in balloons in Fig. 1. Measurement results may contain errors, the impact of which on the sacrificed bits can be estimated by calculating derivatives of the leakage information and its variance given in a reference. Security threats have been discussed in section II-1-4. At the heart of the countermeasures is the minimization of information for the basis selection and bit value. We here focus on characterization; other important issues such as random number generation and side channel vulnerability are briefly discussed in section II-1-4.

*State Preparation (Transmitter)*

QKD systems send random numbers encoded as nonorthogonal photon states from Alice to Bob. The BB84 protocol, for example, employs four states; i.e., two orthogonal states in two complementary bases. The photons should be indistinguishable for Eve not to be given any information. Therefore, the transmitter should emit

- correct states and
- photon pulses identical to those for states used for encoding in terms of timing, shape, spectrum, and other characteristics.

The former issue is related to calibration and is discussed later. Characterization of the output pulses is necessary to ensure the identity. We can set criteria for the identity of the pulses. For example, spectral resolution is limited by the pulse duration, and the distinguishability of the two states in the spectrum region can be characterized by cross-correlation convoluted with the resolution. Similar criteria can be set for the timing and pulse shape.

It may be difficult to ensure the similarity of the pulses if independent lasers are used to represent different quantum states. The difficulty with using lasers comes from the fact that they are active nonlinear devices. The rise time and intensity of the laser pulses will differ according to the bit sequence because of carrier accumulation that is dependent on the bit pattern (the so-called pattern effect). This effect becomes significant as the pulse frequency increases. It would be safer to design the transmitter using a single laser for the light source. In this design, the laser is driven with fixed periodic pulses so that there is no pattern effect. The photon states are prepared with a modulator, whose properties should be independent of the output states. Since the modulators operate linearly, it is easier to satisfy the conditions.

The decoy method requires precise control of the pulse intensity (average photon number). The attenuation of a fast modulator, such as a lithium niobate modulator, may drift and one needs to monitor it to adjust the designed values. This can be done using a combination of a fast modulator and a fixed stable attenuator. The output of the fast modulator is branched and monitored at a test point by a photodetector. The power-controlled output of the fast modulator is further attenuated by the fixed attenuator to a designed average photon number. In practical systems, the accuracy of the power is finite. Furthermore, it is difficult to control the intensity of each pulse. We need to be satisfied with measuring a distribution of the pulse intensity. It is necessary to develop a theory to estimate the number of sacrificed bits in the framework of a finite length code. However, if the distribution is sharp enough, the effect of the power accuracy may be negligible.

The decoy method also assumes a Poisson distribution for the photon number in the transmitter. The photon number distribution should be rigorously confirmed. However, we may safely assume a Poisson distribution because the system employs a laser light source with heavy attenuation.

*Measurement (Receiver)*

The receiver should also detect photons with the same detection efficiency for all expected photon states, otherwise Eve may make use of the tendency of detecting a specific state to obtain information on the states sent. Analysis [Lo et al., 2005a; Hayashi, 2009] suggests that the detection efficiencies may depend on the basis, as long as the detection efficiencies for 0 and 1 are identical using each basis. It has been shown that security can be proved for appropriately characterized detectors with detection efficiencies for 0 and 1 that are mismatched [Fung et al., 2008]. Further theoretical study is needed to apply practical situations.

Note that the afterpulse effect may provide another loophole; the effect causes correlation between the photon detection events because it increases the probability that the same detector fires with successive gate pulses. It is necessary to measure the afterpulse probability as a function of the gate interval, and use a sufficient blanking time for the afterpulse effect to be reduced.

### II – 3 – 3 Trusted Device

One may improve the final key rate by setting the assumption that Eve can access the apparatuses of Alice and Bob only through the transmission channel. The assumption eliminates the effects of errors originating from the apparatus, such as imperfections in the encoder and decoder and dark counts for the detectors. It is, however, necessary to measure such imperfections precisely and to ensure that the measured values are free from Eve's actions. Since we assume that Eve can fully control the quantum channel, a reliable method is to restrict the measurement to a local measurement. Bob can measure the dark counts locally simply by closing the input of the receiver. On the other hand, Bob (or Alice) needs a calibration standard for the decoder/encoder at the local site to measure the residual imperfection. It is safer to use only the dark counts as a local error source if it is difficult to prepare reliable standards. In this case, we regard errors from the imperfections in the encoder and decoder as the result of eavesdropping.

### II – 3 – 4 Calibration and Drift

Alice and Bob need to match their bases and amount of modulation before quantum communication. If Alice and Bob are together in a laboratory, the procedure for this initial calibration is almost trivial (though the calibration itself is not a trivial task.) They can calibrate their devices by connecting directly and measuring the interference with a strong light. However, if they are apart, the calibration is no longer simple because Eve may control the channel. She can influence the measurement results and drive the wrongly calibrated devices. If a strong light is used for calibration, Eve can obtain full information on the state of the light and control the measurement result for each state. Though it is an open question as to how Eve can gain information under the above condition, calibration should be done locally. If single-photon states are used, Eve's strategy is restricted to being state-independent. Her operation is then restricted to unitary transforms and is harmless.

It is common that an apparatus is calibrated at the beginning of communication and this setting is kept through the working period. The idea of initial calibration assumes implicitly that the state of the apparatus is stable during quantum communication once the calibration is established. However, the apparatus state may change over a long time scale. In phase coding, for example, the phase shift in the modulators gradually changes for the same applied signal. The path length difference in the interferometer shifts according to the temperature. As a result, the photon state may wander over the Bloch sphere. Calibration is thus necessary during communication to compensate for the drift before the error of the devices become significant. Therefore, the devices should be stable so as to keep the period between the calibrations practical.

Another issue arises from the drift and fluctuation of the channel. The security analysis assumes that the eavesdropping strategy is independent of time, so that we collect as large a number of events (data) as possible to improve the estimation of the channel properties or leakage of information to Eve. If the channel properties significantly change during data collection, the variance in the distribution increases with the cost of a large number of sacrificed bits. If we know of the absence of Eve, we can measure the drift and compensate for it (in this case we do not need cryptographic protocols at all.) However, if we assume Eve is in the channel, we need to consider that the measurement results are under her control. In practice, the drift of the optical fiber would be slow enough to collect sufficient data. This may become an issue for quantum communication between a low Earth-orbit satellite and the Earth. Transmission loss would change by 10 dB over a few minutes. The transmittance and relative phase would change with fluctuations in the position and direction of the satellite, fluctuations in the refractive index of the air (due to wind), and scattering by small particles. We need to develop a method for the efficient estimation of channel properties using only a small number of data.

### II – 3 – 5 Synchronization

Besides the security proof, it is necessary to compensate fluctuations in an optical fiber set outside the system to achieve continuous operation. Environmental temperature changes, wind, and other environmental conditions can cause large fluctuations in the properties of optical fibers outside the system. The system should be equipped with bit/frame synchronization technologies, fault detection via the monitoring of the QBER, and a resynchronization mechanism. To modulate and detect photon signals with the correct timing, we transmit bit-synchronization signals using wavelength division multiplex (WDM) technology. Since the bit-synchronization signal power is of ordinarily level, spontaneous emissions of laser diodes and nonlinear crosstalk from synchronization signals become a problem. We suppress the crosstalk using WDM filters and a synchronization signal power control. We compensate for the group velocity difference (GVD), which depends on the transmission line, between the quantum signal and the synchronization using an adaptive GVD compensation mechanism. Note that the clock and frame synchronizations do not provide any new information to Eve.

Frame synchronization is indispensable for basis reconciliation and the following key distillation procedures. However, not all the source key bits that are launched from Alice reach Bob in the single-photon transmission, so we cannot apply conventional fixed pattern matching frame synchronizations such as those for SONET/SDH or Ethernet networks. Therefore, we introduce a flame pulse signal that is transmitted with a bit-synchronization signal using WDM technology and a QBER calculation method. The frame position is adjusted roughly by using the frame pulse signal and precisely by using the QBER calculation in the basis reconciliation procedure. We can establish frame synchronization by shifting the bit correspondence while monitoring the QBER until it is much less than 50%. In actual use, even if stable operation is established, we must prepare for accidental faults such as performance degradation caused by synchronization loss or eavesdropping. We introduce fault detection and distinction by QBER monitoring. We also introduce a resynchronization mechanism corresponding to the fault. When bit synchronization worsens, the QBER slowly degrades, and when frame synchronization loss occurs, the QBER rapidly degrades to over 50%. If the QBER degradation is a factor of the bit phase shift or frame phase shift, it would improve after resynchronization. If it does not improve, the fault is classified as a fatal error, which includes that caused by eavesdropping, and the key distribution is aborted.

# Part III: Toward New Generation Quantum Cryptography

It is generally said that there are at least two deep valleys in a route from invention to innovation. One is the Valley of Death between basic research and applied research. The other is the Darwinian Sea, in which industrial species need to be able to supply viable products and services to survive. QCT may have already crossed over the Valley of Death and is now in the Darwinian Sea. QCT should eventually be appreciated by those who have no knowledge of quantum theory. In the early stage of the Struggle for Life in the Darwinian Sea, a crucial issue will be how to embed QCT into photonic networks that are the basic infrastructure of the Internet society. A short-term strategy is to combine current QCT and photonic network technology with reasonable assumptions on nodes and compromises of the security level. The long-term strategy is to invent new schemes that enjoy the merits of all known protocols and to study and develop the quantum repeater to realize full quantum networking. The new paradigm may be referred to as New Generation Quantum Cryptography. The purpose of this part of the report is to discuss briefly the progression from short-term to long-term strategies toward New Generation Quantum Cryptography.

### III – 1 Photonic Network: a Post-IP Network

The total number of digital subscriber line, cable, and FTTH subscribers to the Nippon Telegraph and Telephone Corporation (NTT) Group in Japan reached almost 28 million by the end of September 2007. Of this total, the number of FTTH subscribers exceeded 10 million. Anticipating great changes, the NTT Group announced its Medium-Term Management Strategy in November 2004, pushing ahead with the construction of a NXGN (Next Generation Network). The NXGN will have an important part to play in society as a reliable social infrastructure. Today in the field of telecommunications, safety and security concerns extend from ensuring secure communications to the protection of personal information [Miura, 2008].

At the frontier of optical communications, more advanced technologies are being exploited to realize a sustainable info-communication society that is safe and secure, even beyond the NXGN. The advanced technology of a photonic network adopts

- multi-ary modulations of the amplitude and phase of light,
- all-optical processing of signals, and
- advanced optical switching.

Photonic network technology is expected to serve as a platform for the post-NXGN, which is referred to as the NWGN (New Generation Network). Various applications such as real-time video streaming, multipoint video communication, grid computing, digital cinema, sensor networks, and network games will be provided for by the seamless control of QoS on a photonic transport platform. The key technologies of the photonic transport platform consist of data-granularity-adaptive multi-QoS photonic transport, one-hop transparent links, and autonomously controlled power-minimum photonic networks.

A paradigm of the NWGN consists of a service stratum and transport stratum with a management layer, evolving from legacy network architecture, and is called Open Systems Interconnection (OSI). OSI is standardized by ISO 7498 and consists of seven layers: from the bottom, the physical, data, network, transport, session, presentation, and application layers. There is no doubt that advanced optical switching and optical transmission will play more important roles in the transport layer of the NWGN. An urgent issue is to study how to ensure information security in such a network. This includes the protection of interests of those relying on information systems from failures of confidentiality, integrity, and availability [OECD].

QCT is one of the most promising solutions; however, its current performance needs to be dramatically improved for it to be applied to the NWGN, which requires a breakthrough technology. A practical solution should consist of diverse methods of combining photonic and quantum technologies to realize security in the physical and data layers. This is referred to as Secure Photonic Network technology.

### III – 2 How to embed QCT into Photonic Networks

The security services and their mechanisms in the physical and data layers are summarized as follows [ISO 7498].

(1) Connection Confidentiality: the confidentiality of all user data on a connection.

(2) Connectionless confidentiality: the confidentiality of all user data in a single connectionless session data unit.

(3) Traffic flow confidentiality: the protection of information that might be derived from the observation of traffic flows; for example, the existence, amount, direction, and frequency of communication.

Total encipherment of the data stream is the principal security mechanism in the physical layer and is provided by means of an encipherment device that operates transparently. The objectives of physical layer protection are to protect the entire physical service data bit stream and to provide traffic flow confidentiality. Another mechanism for enhancing the security in the physical layer is route control that avoids vulnerable paths. Physical layer security mechanisms featuring in the photonic network are the technologies of the optical spread spectrum (optical code division multiplex, CDM) and QNRC (Y00).

A mechanism has been proposed for confidentiality in the photonic network to realize the security service required in upper layers as follows. Client data transported in the router network are protected with symmetric-key cryptography, and the optical signal in the optical path network is protected with data encipherment using an optical CDM or QNRC. The secret key needed for both of these cryptosystems can be distributed by QKD. Physical layer cryptosystems are valid if the physical layer interface is provided to clients of the cipher service. The NTT Group has recently demonstrated field experiments on the virtual private network NetMeeting with Vernam's OTP cipher using a key generated by QKD. The system also has a function that acts against falsifications [Honjo et al., 2008].

### III – 3 New Generation Quantum Cryptography

An appropriate starting point in the next phase of quantum cryptography is to integrate different kinds of quantum cryptoschemes into the test-bed of a photonic network. The schemes most suitable to photonic network technology are QNRC and CV-QKD schemes, both of which simply use the presence of quantum noise inherent in lightwave signals. They can be implemented with current technology and hence could be smoothly embedded into a photonic network although the distance of CV-QKD is limited to a few tens of kilometers owing to noise in the fibers. Such schemes may currently suit short-distance applications. Some sophisticated techniques or protocols need to be developed for longer-distance transmission. On the other hand, QKD schemes based on photon counting, say photon QKDs, still face many difficulties in combining with photonic network technologies. For example, how to cope with issues such as wavelength division multiplexing and routing is not trivial. This in turn offers the possibility of investigating new improved schemes.

We wish to exploit all the advantages of these three schemes. For example, a multi-ary differential phase shifting format has better transmission performance, and it can be combined with photon and homodyne detectors to extract the full potential of the quantum effect. In parallel, we should develop quantum "node" technology, which includes novel photon detectors, quantum repeaters, and quantum signal processing to extend the distance of the network. This would lead us to a New Generation Quantum Cryptography that could realize high-speed and secure communications on a photonic transport platform.

### III – 4 QKD in Space

Free-space quantum cryptography between a ground station and satellite is a possible solution for sending quantum information over distances further than what is possible using optical fibers since there is no birefringence effect in the atmosphere.

The European Space Agency (ESA) plans to demonstrate a quantum key distribution experiment called Space-QUEST onboard the ESA Columbus module onboard the International Space Station. QKD employing entangled photons will be demonstrated using three optical ground stations in Europe. Therefore, there is the possibility for Japan to join the project by providing one of the optical ground stations for the global QKD. The National Institute of Information and Communications Technology (NICT) was invited to participate in the round table meeting held at the ESA and cooperation between the ESA and NICT was agreed for the proposal of Space-QUEST.

In Japan, feasibility studies for space QKD have begun for the purpose of developing onboard space QKD terminals. There are two main protocols for the space QKD: the WCP with a decoy state (BB84) and quantum entanglement (E91). The onboard QKD terminals will have 125- and 100-mm-diameter telescopes. Space QKD will be possible under some conditions. It is most important to consider the potential users for such long-distance QKD in terrestrial QKD networks.

### III – 5 Quantum Repeaters

The main obstacle in QKD is the limitation posed by the difficulty of transporting quantum information between distant nodes mainly caused by a dissipative and noisy channel as part of the photonic network. For quantum communication, we cannot use amplifiers before the signals dissipate in the channel because their operation destroys the original quantum information.

One of the most important achievements in the field of quantum information is the discovery of schemes to overcome the limitation of using practical physical resources. Such a system is called a quantum repeater and it allows long-distance quantum communication over noisy channels. As is similar for classical repeaters, the photonic network channels are divided into several segments. At the ends of each segment, local operations are performed for the nodes. These nodes once 'memorize' the quantum information in QKD and the photonics mediate the 'connection' of these segments. Repeating appropriate operations for the nodes enables the expansion of the connection and transmitting of photonic qubits over long distances.

Quantum repeaters have so far been studied mainly for photon QKD because it is based on a simpler signal format. Several models of quantum repeaters for photon QKD have been proposed and some fundamental demonstrations have been experimentally performed in recent years. For practical application in photon QKD systems, it is desirable to realize a repeater that is a solid-state system, like a semiconductor. In such a system, electron and nuclear spins in the material work as nodes for storing the quantum information. Most semiconductor-based schemes for quantum logic rely on nearest-neighbor spin–spin coupling. At each node, photons and spins couple via waveguide structures in a semiconductor chip. An interface between the photons and spins is now being developed using microphotonic and nanophotonic technologies for high-speed networking.

Some protocols for the repeaters are also under consideration for the nesting of the photonic network. A simple method is to use single-photons emitted from the semiconductor nodes. Because the emitted photons strongly correlate with the internal electron spins in the nodes, they act as mediators, and the interferences enable spin–spin coupling by the remotely separated semiconductor nodes. Another method is to communicate the nodes via coherent state laser pulses. The quantum information of distant electron spins is intermediated through the coherent-state laser pulse, and homodyne detection of the pulses projects the expected spin–spin coupling with high probability. Although some probability of errors is inherent in these protocols, the errors might be removed through quantum purifications protocols also operating at the nodes.

A quantum repeater for CV QKD might be more involved. It requires ensembles with many degrees of freedom to store continuously modulated signals. Such a quantum repeater should also be useful in photonic network technology.

The quantum repeater is essential for the next generation of QKD networks, and several approaches are in trial both experimentally and theoretically. Research and design of the quantum repeater should be an important strategic target in the next ten-year plan.

# Part IV: Summary

In this report, we have presented the starting point for the development of quantum cryptography. The main issues discussed were the standardization of quantum cryptography and the research and design for New Generation Quantum Cryptography.

In Part I, we overviewed the current status of quantum cryptography, including theoretical and experimental achievements of both QKD and the QNRC. The QNRC is ready to be applied, or might have already been applied, to practical networks having a range of a few hundred kilometers. QKD needs further improvement for wider use but already has applications such as a Physically Secure Private Network over a short distance and random number sharing between mobile terminals.

Part II was devoted to standardization issues, which mainly relate to security and interoperability. We first summarized the security specifications and requirements. To this end, most known protocols and schemes were described in terms of a) encoding schemes, b) signal transmission schemes, c) measurement and decoding schemes, d) classical communication, and e) status of the security proof. We then discussed how to define security in view of various levels of attacks and practical imperfections.

We summarized the interoperability specifications and requirements. One is the interoperability between QCT and contemporary cryptographic systems and the other is that among quantum cryptosystems. Most of the former issues are left open for subsequent versions. For the latter, we put an emphasis on how to integrate different schemes into QKD networks. A seminal work of SECOQC is reviewed in this respect.

Issues relating to test requirements were also discussed.

In Part III, we discussed how to combine QCT with photonic network technology, and how to realize Secure Photonic Networks to sustain our future info-communication networks. A short-term strategy is to combine current QCT and photonic network technology with reasonable assumptions for the nodes and a compromise of the security level. The long-term strategy is to invent new schemes that have the merits of all known protocols and study and develop quantum repeaters that can realize full quantum networking. We referred to the new paradigm as New Generation Quantum Cryptography.

The aim of this report was to provide an overview of issues in the efficient introduction of quantum information technology to secure our information technology society.

Our philosophy in writing this report is that maintaining comprehensibility and sustainability is essential in the efficient introduction of new technologies into the real world. Our intention was to promote comprehensibility and sustainability through discussions on standardization and new research and design in particular. With this in mind, we intend to partake in further activities.

1. Continuation of the UQC working group:
   The working group will update the report to our best knowledge, giving the status of arts and trends.

2. Contribution to related standardization activities:
   Updating the report so that it can be put into the existing schemes such as CRYPTREC, CC, and FIPS, we will contribute to related standardization activities via collaboration with modern cryptographic societies.

3. Collaboration with academia:
   Collaborating with academia, we will suggest reasonable R&D plans for the further development of quantum information technology.

We would sincerely appreciate any suggestions on and/or cooperation with our activities.

# References

[Assche, 2006] G. V. Assche, Cambridge University Press (2006)
[Barbosa et al., 2003] G.A. Barbosa, E. Corndorf, P. Kumar, and H. P. Yuen, Phys. Rev. Lett. 90, 227901 (2003).
[Beaudry et al., 2008] N. J. Beaudry, T. Moroder, and N. Lütkenhaus, Phys. Rev. Lett. 101, 093601 (2008).
[Bennett and Brassard, 1984] C.H. Bennett, and G. Brassard, in Proceedings IEEE Int. Conf. on Computers, Systems and Signal Processing, Bangalore, India (IEEE, New York), p. 175 (1984).
[Bennett, 1992] C.H. Bennett, Phys. Rev. Lett. 68, 3121 (1992).
[Bennett et al., 1992a] C.H. Bennett, F. Bessate, G. Brassard, L. Salvail, and J. Smolin, J. Crypt. 5, 3 (1992).
[Bennett et al., 1992b] C.H. Bennett, G. Brassard, and N.D. Mermin, Phys. Rev. Lett. 68, 557 (1992).
[Ben-Or et al., 2005] M. Ben-Or, M. Horodecki, D. W. Leung, D. Mayers, J. Oppenheim, Theory of Cryptography: Second Theory of Cryptography Conference, TCC 2005, J.Kilian (ed.) Springer Verlag 2005, vol. 3378 of Lecture Notes in Computer Science, pp. 386-406.
[Bienfang et al., 2004] J. Bienfang, A. Gross, A. Mink, B. Hershman, A. Nakassis, X. Tang, R. Lu, D. Su, C. Clark, C. Williams, E. Hagley, and J. Wen, Opt. Express 12, 2011 (2004).
[Biham and Mor, 1997] E. Biham, and T. Mor, Phys. Rev. Lett. 78, 2256 (1997).
[Boileau et al., 2005] J.-C. Boileau, K. Tamaki, J. Batuwantudawe, R. Laflamme, J.M. Renes, Phys. Rev. Lett. 94 040503 (2005).
[Brassard et al., 2000] G. Brassard, N. Lütkenhaus, T. Mor, and B.C. Sanders, Phys. Rev. Lett. 85, 1330 (2000).
[Buttler et al., 2000] W.T. Buttler, R. J. Hughes, S. K. Lamoreaux, G. L. Morgan, J. E. Nordholt, and C. G. Peterson, Phys. Rev. Lett. 84, 5652 (2000).
[Cerf et al., 2001] N.J. Cerf, M. Lévy, and G. Van Assche, Phys. Rev. A63, 052311 (2001).
[Cirac et al., 2006] M. M. Wolf, G. Giedke, and J. I. Cirac, Phys. Rev. Lett. 96, 080502 (2006).
[Curty et al., 2004] M. Curty, M. Lewenstein, N. Lütkenhaus, Phys. Rev. Lett. 92, 217903 (2004)
[Curty et al., 2007] M. Curty, L-L Zhang, H-K Lo, and N. Lütkenhaus, QIC Vol 7,p. 665-688 (2007).

[Diamanti et al., 2006] E. Diamanti, H. Takesue, C. Langrock, M. M. Fejer and Y. Yamamoto, Opt. Express 14, 13073 (2006).

[Dianati et al., 2008] M. Dianati, R. Alléaume, M. Gagnaire, X. Shen, Security and Communication Networks 1, 57 (2008).

[Dûsek et al., 2000] M. Dusek, M. Jahma, and N. Lütkenhaus, Phys. Rev. A 62, 022306 (2000).

[Ekert, 1992] A.K. Ekert, Phys. Rev. Lett. 67, 661 (1992).

[Elliott, 2002] C. Elliott, New J. Phys. 4, 46 (2002).

[Elliott et al., 2005] C. Elliott, A. Colvin, D. Pearson, O. Pikalo, J.Schlafer, H.Yeh, Proceedings SPIE The International Society for Optical Engineering 5815, 138 (2005); quant-ph/arXiv:0503058.

[Erven et al., 2008] C. Erven, C. Couteau, R. Laflamme, and G. Weihs, quant-ph/arXiv:08072289.

[Fossier et al., 2008] S. Fossier, E. Diamanti, T. Bebuisschert, A. Villing, R. Tualle-Brouri, P. Grangier, New J. Phys. 11 045023 (2009)..

[Fung et al., 2008] C.-H. F. Fung, K. Tamaki, B. Qi, H.-K. Lo, X. Ma, quant-ph/arXiv:0802.3788.

[Garcia-Patron and Cerf, 2006] R. Garcia-Patron and N. J. Cerf, Phys. Rev. Lett. 97, 190503 (2006).

[Goldenberg and Vaidman, 1995] L. Goldenberg and L. Vaidman, Phys. Rev. Lett. 75, 1239 (1995).

[Gottesman and Lo, 2003] D. Gottesman, and H.-K. Lo, IEEE Transactions on Information Theory 49, 457 (2003).

[Gottesman et al., 2004] D. Gottesman, H.-K. Lo, N. Lütkenhaus, and J. Preskill, Quant. Inf. Comput. 4, 325 (2004).

[Grosshans and Grangier, 2002] F. Grosshans, and P. Grangier, Phys. Rev. Lett. 88, 057902 (2002).

[Grosshans et al., 2003] F. Grosshans, G. van Assche, J. Wenger, R. Brouri, N. J. Cerf, and P. Grangier, Nature 421, 238 (2003).

[Grosshans and Cerf, 2004] F. Grosshans and N. J. Cerf, Phys. Rev. Lett. 92, 047905 (2004).

[Hasegawa et al., 2005] T. Hasegawa, T. Nishioka, H. Ishizuka, J. Abe, K. Shimizu, M. Matsui, European Quantum Electronics Conference (EQEC '05), 304 (2005).

[Hasegawa et al., 2007] J. Hasegawa, M. Hayashi, T. Hiroshuma, A. Tanaka, and A. Tomita, quant-ph/arXiv:07053081.

[Hayashi 2009] M. Hayashi, Phys. Rev. A 79, 020303 (2009).

[Hillery, 2000] M. Hillery, Phys. Rev. A 61, 022309 (2000).

[Hirano et al., 2006] T. Hirano, A. Shimoguchi, K. Shirasaki, S. Tokunaga, A. Furuki,

Y. Kawamoto, and R. Namiki, Proc. SPIE vol. 6244 (2006).
[Hirota et al., 2005] O. Hirota, M. Souma, M. Fuse, and K. Kato, Phys. Rev. A 72, 022335 (2005).
[Hiskett et al., 2006] P.A. Hiskett, D. Rosenberg, C. G. Peterson, R. J. Hughes, S. Nam, A. E. Lita, A. J. Miller and J. E. Nordholt, New J. Phys. 8, 193 (2006).
[Hitachi, 2007] Hitachi Information & Communication Engineering, Ltd., demonstration at Nature Photonics Technology Conference 2007, Tokyo. See also http://www.hitachi-jten.co.jp/products/y00/index.html
[Honjo et. al., 2008] T. Honjo, H. Takesue, H. Kamada, K. Tamaki, H. Shibata, K. Shimizu, Y. Tokura, S. Yamamoto, T. Yamamoto, Y. Nishida, O. Tadanaga, M. Asobe, and K. Inoue, "Field trial of differential-phase-shift QKD using polarization independent frequency up-conversion detectors," SECOQC Conference 2008, Vienna, Oct.10, 2008.
[Horodecki et al., 2005] K. Horodecki, M. Horodecki, P. Horodecki, J. Oppenheim, Phys. Rev. Lett. 94, 160502 (2005).
[Huttner et al., 1995] B. Huttner, N. Imoto, N. Gisin, and T. Mor, Phys. Rev. A51, 1863 (1995).
[Hwang, 2003] W.-Y. Hwang, Phys. Rev. Lett. 91, 057901 (2003).
[idQuantique] http://www.idquantique.com/
[Inamori et al., 2001] H. Inamori, N. Lütkenhaus, D. Mayers, Eur. J. Phys. D 41, 599 (2007), quant-ph/arXive:0107017.
[Inoue et al., 2002] K. Inoue, E. Waks, and Y. Yamamoto, Phys. Rev. Lett. 89, 037902 (2002).
[Inoue et al., 2003] K. Inoue, E. Waks, and Y. Yamamoto, Phys. Rev. A68, 022317 (2003).
[Intallura et al., 2007] P.M. Intallura, M. B. Ward, O. Z. Karimov, Z. L. Yuan, P. See, A. J. Shields, P. Atkinson, and D. A. Ritchie, Appl. Phys. Lett. 91, 161103 (2007).
[ISO 7498] International Standard, ISO 7498: Information processing systems – Open Systems Interconnection – Basic Reference Model – Part2: Security Architecture.
[Jennewein et al., 2000] T. Jennewein, C. Simon, G. Weihs, H. Weinfurter, and A. Zeilinger, Phys. Rev. Lett. 84, 4729 (2000).
[Katz and Lindell, 2007] Introduction to Modern Cryptography: Principles and Protocols (Chapman & Hall/CRC Cryptography and Network Security Series, 2007)
[Koashi and Preskill, 2003] M. Koashi, and J. Preskill, Phys. Rev. Lett. 90, 057902 (2003).
[Koashi, 2004] M. Koashi, Phys. Rev. Lett. 93, 120501 (2004).
[Koashi, 2005] M. Koashi, quant-ph/arXive:0505108.

[Koashi et al., 2008] M. Koashi, Y. Adachi, T. Yamamoto, and N. Imoto, ArXiv:0804.0891 (2008).

[Kraus et al., 2005] B. Kraus, N. Gisin, R. Renner, Phys. Rev. Lett. 95, 080501 (2005).

[Lance et al., 2005] A. M. Lance, T. Symul, V. Sharma, C. Weedbrook, T. C. Ralph, and P. K. Lam, Phys. Rev. Lett. 95, 180503 (2005).

[Leverrier et al., 2008] A. Leverrier, E. Karpov, P. Grangier, and N.J. Cerf, quant-ph/arXiv:0809.2252.

[Leverrier and Grangier, 2008] A. Leverrier, and P. Grangier, quant-ph/arXive:0812.4246.

[Ling et al., 2008] A. Ling, M.P. Peloso, I. Marcikic, V. Scarani, A. Lamas-Linares, and C. Kurtsiefer, Phys. Rev. A 78, 020301 (2008).

[Lo and Chau, 1999] H.-K. Lo and H.F. Chau, Science 283, 2050 (1999).

[Lo et al., 2005a] H.-K. Lo, H.F. Chau, and M. Ardehali, J. Crypt. 18, 133 (2005).

[Lo et al., 2005b] H.-K. Lo, X. Ma, and K. Chen, Phys. Rev. Lett. 94, 230504 (2005).

[Lodewyck et al., 2007] J. Lodewyck, M. Bloch, R. García-Patrón, S. Fossier, E. Karpov, E. Diamanti, T. Debuisschert, N. J. Cerf, R. Tualle-Brouri, S. W. McLaughlin, and P. Grangier, Phys. Rev. A 76, 042305 (2007).

[Lütkenhaus, 2000] N. Lütkenhaus, Phys. Rev. A 61, 052304 (2000).

[MagiQ] http://www.magiqtech.com/

[Mayers, 1996] D. Mayers, Advances in Cryptology — Proceedings of Crypto '96 (Springer Verlag, Berlin), 343 (1996).

[Miura, 2007] S. Miura, NTT Technical Review Vol. 6 No. 4 Apr. (2008), pp.?

ITU-T Recommendation Y.2701 : "Security requirements for NGN release 1, "April (2007).

[Miyazaki et al., 2005] T. Miyazawa, K. Takemoto, Y. Sakuma, S. Hirose, T. Usuki, N. Yokoyama, M. Takatsu and Y. Arakawa, Jpn. J. Appl. Phys. 44, L620 (2005).

[Mo et al., 2005] X.F. Mo, B. Zhu, Z.F. Han, Y.Z. Gui, and G.C. Guo, Optics Lett. 30, 2632 (2005).

[Muller et al., 1996] A. Muller, H. Zbinden and N. Gisin, Europhys. Lett. 33 335 (1996).

[Muller et al., 1997] A. Muller, T. Herzog, B. Huttner, W. Tittel, H. Zbinden, and N. Gisin, Appl. Phys. Lett. 70, 793 (1997).

[Naik et al., 2000] D.S. Naik, C. G. Peterson, A. G. White, A. J. Berglund, and P. G. Kwiat, Phys. Rev. Lett. 84, 4733 (2000).

[Nair et al., 2006] R. Nair, H. P. Yuen, E. Corndorf, T. Eguchi, and P. Kumar, Phys. Rev. A 74, 052309 (2006).

[Nambu et al., 2004] Y. Nambu, T. Hatanaka and K. Nakamura, Jpn. J. Appl. Phys.

43, L1109 (2004).
[Navasques et al., 2006] M. Navasques, F. Grosshans, and A. Acin, Phys. Rev. Lett. 97, 190502 (2006).
[NuCrypt, 2007] NuCrypt, demonstration at Milcom 2007. See also http://www.nucrypt.net/index.html for the latest information.
[OECD] OECD Guidelines for the Security of Information Systems, 1992.
[Peng et al., 2007] C.-Z. Peng, J. Zhang, D. Yang, W.-B. Gao, H.-X. Ma, H. Yin, H.-P. Zeng, T. Yang, X.-B. Wang, and J.-W. Pan, Phys. Rev. Lett. 98, 010505 (2007).
[Phoenix et al., 2000] S. Phoenix, S.M. Barnett and A. Chefles, J. Mod. Opt. 47, 507 (2000).
[Poppe et al., 2008] A. Poppe, M. Peev and O. Maurhart, Int. J. Quantum Information 6, 209 (2008).
[Ralph, 1999] T.C. Ralph, Phys. Rev. A 61, 010303(R) (1999).
[Reid, 2000] M.D. Reid, Phys. Rev. A 62, 062308 (2000).
[Renner and König, 2005] R. Renner, R. König, Second Theory of Cryptography Conference, TCC 2005. Volume 3378 of Lecture Notes in Computer Science., Springer (February 2005) 407–425.
[Renner and Cirac, 2008] R. Renner, and J. I. Cirac, Phys. Rev. Lett. 102, 110504 (2009)..
[Renner and Cirac
[Rosenberg et al., 2007] D. Rosenberg, J. W. Harrington, P. R. Rice, P. A. Hiskett, C. G. Peterson, R. J. Hughes, A. E. Lita, S. W. Nam, and J. E. Nordholt, Phys. Rev. Lett. 98, 010503 (2007).
[Scarani et al., 2004] V. Scarani, A. Aćin, G. Ribordy, and N. Gisin, Phys. Rev. Lett. 92, 057901 (2004).
[Scarani and Renner, 2008], V. Scarani and R. Renner, Phys. Rev. Lett. 100, 200501 (2008).
[Schmitt-Manderbach et al., 2007] T. Schmitt-Manderbach, H. Weier, M. Fürst, R. Ursin, F. Tiefenbacher, T. Scheidl, J. Perdigues, Z. Sodnik, C. Kurtsiefer, J. G. Rarity, A. Zeilinger, and H. Weinfurter, Phys. Rev. Lett. 98, 010504 (2007).
[Shor and Preskill, 2000] P.W. Shor and J. Preskill, Phys. Rev. Lett. 85, 441 (2000).
[SmartQuantum] http://www.smartquantum.com/SmartQuantum.html
[Silberhorn et al., 2002] C. Silberhorn, T. C. Ralph, N. Lütkenhaus, and G. Leuchs, Phys. Rev. Lett. 89, 167901 (2002).
[Soujaeff et al., 2007] A. Soujaeff, T. Nishioka, T. Hasegawa, S. Takeuchi, T. Tsurumaru, K. Sasaki, and M. Matsui, Opt. Express 15, 726 (2007).
[Stucki et al., 2002] D. Stucki, N. Gisin, O. Guinnard, G. Ribordy, and H. Zbinden, New J. Phys. 4, 41 (2002).

[Tajima et al., 2007] A. Tajima, A. Tanaka, W. Maeda, S. Takahashi, A. Tomita, IEEE J. Sel. Topics Quantum Electron. 13, 1031 (2007).

[Takesue et al., 2007] H. Takesue, S. W. Nam, Q. Zhang, R. H. Hadfield, T. Honjo, K. Tamaki and Y. Yamamoto, Nature Photonics 1, 343 (2007).

[Tamaki et al., 2003] K. Tamaki, M. Koashi, and N. Imoto, Phys. Rev. Lett. 90, 167904 (2003).

[Tamaki and Lütkenhaus, 2004] K. Tamaki and N. Lütkenhaus, Phys. Rev. A 69, 032316 (2004).

[Tamaki and Lo, 2006] K. Tamaki and H.-K. Lo, Phys. Rev. A 73, 010302(R) (2006).

[Tamaki et al., 2006] K. Tamaki, N. Lütkenhaus, M. Koashi, and J. Batuwantudawe, quant-ph/arXive:0607082.

[Tanaka et al., 2005] A. Tanaka, W. Maeda, A. Tajima and S. Takahashi, The 18th Annual Meeting of the IEEE Lasers and Electro-Optics Society (LEOS2005), 557 (2005).

[Tanaka et al., 2008] A. Tanaka, M. Fujiwara, S. W. Nam, Y. Nambu, S. Takahashi, W. Maeda, K. Yoshino, S. Miki, B. Baek, Z. Wang, A. Tajima, M. Sasaki, and A. Tomita, Opt. Express 16, 11354 (2008).

[Tittel et al., 2000] W. Tittel, J. Brendel, H. Zbinden, and N. Gisin, Phys. Rev. Lett. 84, 4737 (2000).

[Townsend, 1994] P.D. Townsend, Electron. Lett. 30, 809 (1994).

[Tsurumaru, 2007] T. Tsurumaru, Phys. Rev. A 75, 062319 (2007)

[Tsurumaru and Tamaki, 2008] T. Tsurumaru and K. Tamaki, Phys. Rev. A 78, 032302 (2008).

[Ursin et al., 2007] R. Ursin, F. Tiefenbacher, T. Schmitt-Manderbach, H. Weier, T. Scheidl, M. Lindenthal, B. Blauensteiner, T. Jennewein, J. Perdigues, P. Trojek, B. Ömer, M. Fürst, M. Meyenburg, J. Rarity, Z. Sodnik, C. Barbieri, H. Weinfurter, and A. Zeilinger, Nature Physics 3, 481 (2007)

[Waks et al., 2006] E. Waks, Hiroki Takesue, and Yoshihisa Yamamoto, Phys. Rev. A 73, 012344 (2006).

[Wang, 2005] X.-B. Wang, Phys. Rev. Lett. 94, 230503 (2005).

[Wegman and Carter, 1981] M. N. Wegman and J. L. Carter, J. Comp. Syst. Sci. 22, 265 (1981)

[Wen et al., 2008] K. Wen, K. Tamaki, and Y. Yamamoto, ArXiv:0806.2684 (2008).

[Vernam, 1926] G.S. Vernam, J. AIEE 45, 109, (1926).

[Yuan and Shields, 2005] Z.L. Yuan and A.J. Shields, Opt. Express 13, 660 (2005).

[Yuan et al., 2007] Z.L. Yuan, A.W. Sharpe, and A.J. Shields, Appl. Phys. Lett. 90, 011118 (2007).

[Yuan et al., 2008] Z.L. Yuan, A.R. Dixon, J.F. Dynes, A.W. Sharpe, A.J. Shields,

Appl. Phys. Lett. 92, 201104 (2008).
[Yuen, 2003] H. P. Yuen, quant-ph/arXive:0311061.
[Zhao et al., 2006] Y. Zhao, B. Qi, X. Ma, H.-K. Lo, and L. Qian, Phys. Rev. Lett. 96, 070502 (2006).
[Zhao et al., 2008] Y.-B. Zhao, Z.-F. Han, and G.-C. Guo, quant-ph/arXiv:0809.2683.